\pdfoutput=1
\documentclass[12pt, oneside]{article}   	% use "amsart" instead of "article" for AMS\TVX format

\usepackage[svgnames]{xcolor}
\usepackage[colorlinks]{hyperref}

\usepackage[margin=1.25in]{geometry}                		% See geometry.pdf to learn the layout options. There are lots.
\usepackage{graphicx}				% Use pdf, png, jpg, or eps with pdflatex; use eps in DVI mode
\usepackage{caption} 
\usepackage{amssymb}
\usepackage{bm}
\usepackage{amsmath}
\usepackage{bbm}
\usepackage{amsthm}	
\usepackage{natbib}
\usepackage{tikz}
\usepackage{capt-of}
\usepackage{extarrows}
\usepackage{multirow}
\usepackage{booktabs}

\hypersetup{
	citecolor=DarkBlue,
	bookmarksnumbered=true,
	urlcolor=Indigo,linkcolor=blue
}

\setcitestyle{authoryear,open={(},close={)}}

\newtheorem{thm}{Theorem}
\newtheorem{prop}{Proposition}
\newtheorem{ass}{Assumption}

\newtheorem{lemma}{Lemma}

\usepackage{setspace}
\newcommand\independent{\protect\mathpalette{\protect\independenT}{\perp}}
\def\independenT#1#2{\mathrel{\rlap{$#1#2$}\mkern5mu{#1#2}}}
\def\sign{\mathop{\rm sign}}

\title{Testing and Identifying Substitution and Complementarity Patterns\thanks{I am grateful to Sung Jae Jun, Joris Pinkse, and Keisuke Hirano for their invaluable guidance and support. I thank Andres Aradillas-Lopez, Jason Blevins, Jeffrey Campbell, Liyu Dou, Paul Grieco, Patrik Guggenberger, Marc Henry, Ming Li, Xiao Lin, Xun Lu, John Rehbeck, Xiaoxia Shi, Ewout Verriest, Jiahui Xu, Kaida Zhang, and participants at various conferences and seminars.}}
\author{Rui Wang%
\thanks{Department of Economics, The Ohio State University.  Email: \texttt{wang.16498@osu.edu}. }
}
\date{February 27, 2023}

\begin{document}
\maketitle

\begin{abstract}

This paper studies semiparametric identification of substitution and complementarity patterns between two goods using a panel multinomial choice model with bundles. The model allows the two goods to be either substitutes or complements and admits heterogeneous complementarity through observed characteristics. I first provide testable implications for the complementarity relationship between goods. I then characterize the sharp identified set for the model parameters and provide sufficient conditions for point identification. The identification analysis accommodates endogenous covariates through flexible dependence structures between observed characteristics and fixed effects while placing no distributional assumptions on unobserved preference shocks. My method is shown to perform more robustly than the parametric method through Monte Carlo simulations. As an extension, I allow for unobserved heterogeneity in the complementarity, investigate scenarios involving more than two goods, and study a class of nonseparable utility functions. 

%As an empirical illustration, I apply my method to estimate substitution patterns between cigarettes and e-cigarettes using the Nielsen data.

\hspace{8pt} 


%\textbf{JEL}: C14, C33, C35

\end{abstract}

\newpage

\section{Introduction}

Substitution/complementarity relationships between goods have been studied in  various applications such as online news versus print newspapers, digital books versus traditional books, and cigarettes versus e-cigarettes. The relationship plays a crucial role in consumers' decisions; therefore, understanding substitution patterns is important for predicting demand for a good and analyzing the welfare effects of, for example, a merger of two companies or the introduction of a new good \citep*{petrin2002, goolsbee2004, gentzkow2007}. 

The standard multinomial choice models typically assume that consumers can buy only one good at a time, which  rules out complementarity  by assumption. However, even some goods traditionally perceived as substitutes are shown to be complements in different contexts. For example,  \cite{zhao2019} suggests that cigarettes and e-cigarettes could be  complements, and \cite{grzybowski2008} demonstrate the complementarity between telephone calls and messages.

With these motivations in mind, I propose a semiparametric panel multinomial choice model with fixed effects to study substitution/complementarity patterns. This model allows consumers to purchase two goods simultaneously, accommodating the possibility that the two goods are either substitutes or complements. The model also permits heterogeneous complementarity relationships through observed characteristics. 

Identifying substitution/complementarity patterns with bundles involves several challenges. First, the demand for one good includes consumers who buy this good alone and those who buy a bundle. Therefore, a large demand for one good could come from consumers' high utility for this good, or its complementarity with another good, or both.
We need to disentangle the two sources to identify the complementarity relationship. Second, the purchase of two goods together may be due to either the goods' complementarity or the unobserved correlation between consumers' preferences over the two goods. For example, consumers may buy a variety of organic goods because of their preferences over organic goods instead of the complementarity between these goods.
Distinguishing the complementarity relationship and the correlation between consumers' tastes for goods could be challenging since they are both unobserved and can affect consumers' decisions simultaneously.

To tackle these challenges, my paper exploits a conditional stationarity assumption about preference shocks over time, which enables us to use intertemporal variation in conditional choice probabilities for identification. I first provide an approach to test the substitution/complementarity relationship between goods and then derive the sharp identified set for the model parameters. 

The testing approach exploits the relationship between the demand for two goods with the covariate indices of those goods under a substitution or complementarity relationship between goods. This approach does not require the estimation of any model parameter. Instead, it constructs conditional moment inequalities that only depend on observed variables. 
Therefore, testing complementarity can be conducted by directly testing these moment inequalities.

To derive the sharp identified set, the paper uses intertemporal comparisons of conditional choice probabilities to obtain identifying restrictions on the model parameters. The result not only identifies the sign of the complementarity but also bound its value. The sharpness of the identification results is established, indicating that all available information from the data for the model parameters has been exhausted.   Additionally, the paper shows that under large support conditions for the covariates and a linear specification of the complementarity, point identification can be achieved.

For estimation and inference, I characterize conditional moment inequalities based on the identification results and apply the approach in \cite*{shi2018}. In Monte Carlo simulations, I compare the finite sample performance of the method in this paper to that of a parametric method, which assumes a parametric distribution over the error terms and a linear model for fixed effects. The simulation results show that the semiparametric method in this paper has the advantage of performing robustly over different DGP designs, whereas the parametric estimator is vulnerable to misspecifications.

% I characterize conditional moment inequalities based on the identification results and apply a similar approach to \cite*{shi2018} for estimation.
% In Monte Carlo simulations, I compare the finite sample performance of the method in this paper to that of a parametric method that assumes a parametric distribution over the error terms and a linear model for fixed effects. The simulation results show that the semiparametric method in this paper has the advantage of performing robustly over different DGP designs, whereas the parametric estimator performs poorly if either the parametric distribution or the linear model is incorrectly specified.

%As an empirical illustration of this approach, I estimate the substitution pattern between cigarettes and e-cigarettes using the Nielsen Retail Scanner data. The data contain weekly store-level information about the prices and sales of cigarettes and e-cigarettes. The substitution pattern between the two goods is identified from comparisons of the conditional demand for the two goods over different weeks. The estimation results based on this data show that cigarettes and e-cigarettes are substitutes on average.

I explore several extensions to the model. 
 Firstly, I consider the possibility of unobserved heterogeneity in the complementarity relationship, which allows for differences in complementarity across individuals that are not captured by covariates. By exploiting variation in demand for two goods, I characterize partial identification results for the fraction of people for whom the two goods are complements. In the online Appendix, I also extend the model to include more than two goods and establish partial identification results for model parameters under additional assumptions. Additionally, I extend my approach to incorporate nonseparable utility functions as well as cross-sectional models.
 
%I consider multiple extensions of the paper. First, I allow for unobserved heterogeneity in the complementarity so that the complementarity relationship can be different across individuals regardless of covariates. By exploiting variation in demand for two goods, I characterize partial identification results for the fraction of people for whom the two goods are complements. In the online Appendix, I extend the model to more than two goods and establish partial identification results for model parameters under additional assumptions. I also extend my approach to incorporate nonseparable utility functions, as well as cross-sectional models.

%As an extension, I allow for unobserved heterogeneity in the complementarity so that the complementarity relationship can be different across individuals regardless of covariates. Assuming that the distribution of the complementarity is stationarity over time, I characterize partial identification results  for the fraction of people for whom the two goods are complements by exploiting  intertemporal  variation in covariates and conditional demand for two goods. 
%Moreover, I study the case of more than two goods and establish partial identification results for model parameters under some additional assumptions.

 \subsection{Related Literature}
This paper contributes to the literature studying substitution/complementarity patterns in a discrete choice model with bundles.  \cite{gentzkow2007} develops a discrete choice model that allows for complementarity to study the substitution relationship between online news and print newspapers. His paper is flexible about substitution patterns and also admits correlations between error terms across choices. Compared to \cite{gentzkow2007}, my paper does not reply on parametric assumptions, and also allows for flexible dependence structures between covariates and fixed effects.
 \cite*{dunker2015} and \cite{iaria2020} allow for endogeneity and provide identification results of models with bundles by extending the classic BLP approach in \cite*{berry1995}. Their methods use demand inversion and parametric distributions over error terms, and they address endogeneity using instrumental variables. \cite{monardo2021} studies a more flexible model for the inverse demand and uses an instrument to construct moment conditions. My paper mainly exploits intertemporal variation in panel data to derive sharp identification results and the method does not require demand inversion  or instrumental variables. 

%My work allows for unknown distributions of both fixed effects and error terms so it does not specify a parametric model for the demand function (or inverse of the demand). Also, 

 %my paper studies a class of nonseparable and potentially unknown utility functions in an extension of this paper.
% then the methods in the literature do not apply in this scenario.  \cite{fan2003} also allows for the purchase of two goods but assumes the substitutability between the two goods.

There are several papers that allow for unknown distributions of error terms to study substitution patterns. \cite{fox2017} study semiparametric identification of a discrete choice model with bundles under a large support assumption and exogenous covariates.  \cite{allen2022} allow for unobserved heterogeneity in the complementarity and provide partial identification for the fraction of people for whom the two goods are complements. My paper focuses on heterogeneous complementarity through observed covariates and provide sharp identification with bounded support of covariates.  In an extension of this paper, I allow for unobserved heterogeneous complementarity while relaxing the exogenous covariates assumption and the exclusion restriction in \cite{allen2022}. 
%The method can identify the sign and bound the value of the complementarity given consumers' characteristics by exploiting panel data.

My paper is also related to a large body of literature on panel multinomial choice models with fixed effects. \cite{chamberlain1980} provides a conditional fixed effect logit estimator for the panel multinomial choice model under a logistic distribution over disturbances. \cite{manski1987} as well as \cite{honore2002} relax the logistic distribution assumption and study semiparametric identification of a binary choice model. \cite{manski1987} uses a maximum score approach that relies on a group stationarity assumption, and \cite{honore2002} exploit the idea of a special regressor to identify the panel binary choice model. \cite{honore2020}, \cite*{khan2020}, and \cite*{dobronyi2021} study identification of dynamic binary choice models.
 
\cite{pakes2019} and \citet*{shi2018} study a panel multinomial choice model. \cite{pakes2019} derive sharp identification of the model by characterizing conditional moment inequalities, while \citet*{shi2018} use cyclic monotonicity for identification and estimation. \cite*{khan2021} provide inference methods for multinomial choice models.  \cite{gao2020} relax the separable utility function assumption in the previous papers and study a class of nonseparable utility functions. %that can incorporate infinite-dimensional unobserved heterogeneity.
 The aforementioned papers focus on the identification of own price coefficients rather than substitution patterns between different goods. My paper builds on this literature to allow for bundles in the panel multinomial choice model and characterizes sharp identification for substitution patterns.
 
The rest of this paper is organized as follows. Section \ref{sec:base} introduces the panel multinomial choice model with bundles. Section \ref{subsec:test} provides testable implications for the complementarity between two goods. Section \ref{sec:inde} characterizes the sharp identified set for the model parameters and provides sufficient conditions for point identification. Section \ref{sec:esti} develops conditional moment inequalities, and Section \ref{sec:simu} examines the finite sample performance via Monte Carlo simulations. Section \ref{sec:exte} studies an extension. Section \ref{sec:conc} concludes. 
 
%Section \ref{sec:empi} studies substitution patterns between cigarettes and e-cigarettes as an empirical illustration. 

\section{Panel Multinomial Choice Model}\label{sec:base}

This section presents a panel multinomial choice model allowing for bundles. Consider a short-panel structure: let $i\in \mathcal{I}$ denote consumers and $t\leq T$ denote time periods where the length of the panel $T\geq 2$ is fixed. 
Since this paper focuses on substitution patterns between two goods, I consider the case of two goods: $\{A,B\}$.\footnote{Appendix \ref{sec:more} studies the case of more than two goods and establishes partial identification results under additional assumptions.} Instead of assuming that consumers can buy either only good $A$ or only good $B$, this model allows consumers to purchase goods $A$ and $B$ simultaneously and focuses on identifying the complementarity between goods. 
%The possibility of buying the two goods together allows the two goods to be either substitutes or complements. 
 %\footnote{Section \ref{sec:cross} shows that the approach can be applied to cross-sectional models.}
 
The choice set for consumers is $\mathcal{C}=\{A, B, AB, O\}$, where $A$ (or $B$) denotes purchasing only good $A$ (or $B$), $AB$ denotes purchasing $A$ and $B$ simultaneously, and $O$ denotes the outside option. I assume that consumers buy at most one unit of each good, and they select the choice yielding the highest utility in their choice set.

Let $u_{ijt}$ denote consumer $i$'s utility from consuming choice $j\in \mathcal{C}$ at time $t$. Following \cite{gentzkow2007}, let $\Gamma_{it}=(u_{iABt}-u_{iAt})-(u_{iBt}-u_{iOt})$ denote the incremental utility of consuming good $B$ when good $A$ is also consumed. The utility $u_{ijt}$ is specified as\footnote{In Appendix \ref{sec:dis}, I discuss how this utility specification connects to a conventional discrete choice model. In Appendix \ref{subsec:nonsep}, I also investigate a class of nonseparable utility functions.}
%In Appendix \ref{sec:dis}, I discuss how this utility specification connects to treating the bundle as an additional good in a conventional discrete choice model. 
%Let $\Gamma_{it}$ denote the incremental utility from consuming the bundle $AB$ compared to the sum of utilities of consuming goods $A$ and $B$ alone. The utility $u_{ijt}$ of consumer $i$ from consuming choice $j\in \mathcal{C}$ at time $t$ is specified as
\begin{equation} \label{model:u1}
\begin{aligned}
&u_{iAt}=X_{iAt}'\beta_0+\alpha_{iA}+\epsilon_{iAt}, \\
&u_{iBt}=X_{iBt}'\beta_0+\alpha_{iB}+\epsilon_{iBt}, \\
&u_{iABt}=u_{iAt}+u_{iBt}+\Gamma_{it}, \\
&u_{iOt}=0,
\end{aligned}
\end{equation}
Here $X_{ijt}\in \mathbbm{R}^{d_x}$ denotes a vector of observed characteristics, which may include consumer $i$'s characteristics (e.g., income), product $j$'s characteristics (e.g., price), and the interaction terms between them; $\alpha_{ij}\in \mathbbm{R}$ denotes an unobserved individual-specific fixed effect for product $j$ that does not change over time, such as consumers' loyalty to a brand; $\epsilon_{ijt}\in \mathbbm{R}$ denotes an unobserved and time-varying shock that affects consumers' utility over time; and $\beta_0\in \mathbbm{R}^{d_x}$ is a finite-dimensional unknown parameter vector. Without loss of generality, we normalize the utility of the outside option to zero so the utilities of the remaining choices are defined relative to the utility of the outside option.

The sign of the incremental utility $\Gamma_{it}$ represents the complementarity relationship between the goods. Two goods are considered complements if their combined utility is greater than the sum of their individual utilities ($\Gamma_{it}> 0$), and substitutes if the reverse is true ($\Gamma_{it}< 0$).
This definition is equivalent to an alternative definition of substitution patterns using aggregate demand, as discussed in Section \ref{subsec:sub}. In this model, $\Gamma_{it}$ can be either positive, negative, or zero, which allows for the possibility that two goods can be either substitutes or complements. I provide identification results for both the coefficient $\beta_0$ and the complementarity $\Gamma_{it}$.

In addition to the covariate $X_{ijt}$, I assume that consumer $i$'s choice at time $t$ is observed which is denoted as $Y_{it}\in \mathcal{C}$. Consumers select the choice with the highest utility, implying
%this has the following implication:
\begin{equation*}
Y_{it}=j \  \Longrightarrow  \ u_{ijt}\geq u_{ikt} \ \text{for all} \ k\in \mathcal{C}.
\end{equation*}

To simplify analysis, I assume that tie outcomes between choices happen with a zero probability, eliminating the need to account for such situations. This assumption is satisfied as long as one of the error terms is continuously distributed. Even if ties happen with nonzero probability, all identification results hold as long as the tie-breaking rule is fixed over time.
 
%Some examples include that consumers randomly select a choice with a fixed probability, or consumers always pick one specific choice in the same order when tie outcomes happen which is used in \cite{pakes2019}.

The main objective of this paper is to identify the substitution/complementarity relationship and the coefficient $\beta_0$ from consumers' choices $Y_{it}$ and covariates $X_{ijt}$.

Next, I introduce some assumptions about the model.

\begin{ass}\label{ass:par}
The incremental term $\Gamma_{it}$ in the utility $u_{iABt}$ is specified as
\begin{equation*}\label{ass1}
\Gamma_{it}=\Gamma(Z_{i}),
\end{equation*}
%where the function $\Gamma$ is known up to a finite-dimensional parameter $\gamma_0$, and $Z_{i}\in \mathbbm{R}^{d_z}$ denotes a vector of observed characteristics. 
where $Z_{i}\in \mathbbm{R}^{d_z}$ denotes a vector of observed characteristics, and $\Gamma$ can be an unknown function of $Z_i$.  
\end{ass}
 
%Assumption \ref{ass:par} requires the complementarity to depend on observed covariates $Z_{i}$ in a parametric function. For the identification analysis, the parametric form can be relaxed and we can allow for an unknown function $\Gamma$. I use the parametric function to keep a uniform structure for identification and estimation analyses.  The function $\Gamma$ is flexible and can be nonlinear in covariate $Z_{i}$, which would admit rich complementarity patterns. The covariate $Z_{i}$ may include consumers' characteristics such as income and age so that Assumption \ref{ass:par} allows for heterogeneous complementarity relationships through observed characteristics.
% For simplicity of notation, I consider covariate $Z_{i}$ to be fixed over time for any $t$. If the covariate changes over time, then the analysis can be conducted conditional on the same value of the covariate over time: $Z_{is}=Z_{it}=z$.
Assumption \ref{ass:par} requires the complementarity to only depend on observed covariates $Z_{i}$. The function $\Gamma$ is flexible and can be unknown to researchers, which could admit rich complementarity patterns. The covariate $Z_{i}$ may include consumers' characteristics such as income and age so that Assumption \ref{ass:par} allows for heterogeneous complementarity relationships through observed characteristics.
I assume the covariate $Z_{i}$ to be fixed over time for any $t$. If the covariate changes over time, then the same analysis can be conducted conditional on the same value of the covariate over time: $Z_{is}=Z_{it}=z$.

One restriction of Assumption \ref{ass:par} is that it excludes unobserved heterogeneity in the complementarity and assumes the same complementarity relationship for individuals with the same characteristic $Z_i=z$. This structure allows us to not only identify the sign of the complementarity $\Gamma(z)$ but also bound the magnitude of the complementarity $\Gamma(z)$ given $Z_i=z$. 
In Section \ref{sec:exte}, I consider an extension that allows for unobserved heterogeneity in the complementarity $\Gamma_{it}$, where we can only obtain weaker results and partially identify the distribution of the sign of $\Gamma_{it}$.  

%discuss the case in which $\Gamma_{it}$ is a random variable so that it incorporates unobserved heterogeneity into the complementarity relationship.

%Under Assumption \ref{ass:par}, I establish more informative results that not only identify the sign of the complementarity relationship but also bound the magnitude of the complementarity $\Gamma(z)$ given $Z_i=z$. %, such as the effect of introducing a new good. 

\begin{ass}[Exclusion]\label{exc}
There exists at least one characteristic $X^{*}_{it}$ in $X_{it}=(X_{iAt}, X_{iBt})$ that is not in $Z_{i}$, and its coefficient is nonzero.
\end{ass}
The exclusion assumption requires that there exists one variable that only influences the utility for good $A$ or $B$ but not the complementarity between the two goods. One example of this variable is the price of good $A$ or $B$, which affects the utility of a single good but may not influence the complementarity between the two goods. The sign of the coefficient for $X^{*}_{it}$ can still be unknown to researchers. Moreover, this assumption does not restrict the covariate $Z_i$; any variable affecting the complementarity is allowed to influence the utility of a single good.

%The exclusion restriction is a common assumption for identification. Since my model allows for multiple choices,  there are two key components affecting the demand for a good: one is variation in characteristics of a single good and the other is the complementarity relationship between two goods. The exclusion restriction can help distinguish the two sources and identify the substitution patterns. %Without it, it is still possible to establish identification for model \eqref{model:u1} but the results will be less informative. 

The last assumption is the stationarity condition for the distribution of the unobserved shocks. 
Let $X_{it}=(X_{iAt}, X_{iBt}), \alpha_i=(\alpha_{iA}, \alpha_{iB})$, and $\epsilon_{it}=(\epsilon_{iAt}, \epsilon_{iBt})$ collect covariates, fixed effects, and error terms of the two goods. 

\begin{ass}(Stationarity)\label{ass:sta}
The distribution of $\epsilon_{it}$ conditional on $(X_{is}, X_{it}, Z_{i}, \alpha_i)$ is stationary over time; that is,
\begin{equation*}
\epsilon_{is}\mid X_{is}, X_{it}, Z_{i}, \alpha_i   \stackrel{d}{\sim}   \epsilon_{it}\mid X_{is}, X_{it}, Z_{i}, \alpha_i \quad \text{for any} \ s, t\leq T.
\end{equation*}
\end{ass}

This assumption is a multinomial extension of the conditional homogeneity assumption in \cite{manski1987}. It is commonly used in the literature on panel multinomial choice models, including \cite{pakes2019} and \cite*{shi2018}, who study identification of the coefficient $\beta_0$ under this assumption.
Assumption \ref{ass:sta} restricts the conditional distribution of $\epsilon_{it}$ to be stationary over time, but it allows the error term $\epsilon_{it}$ to be dependent across choices and over time. In addition, it does not impose any distributional restrictions on the unobserved term $\epsilon_{it}$. Therefore, the standard logit/probit models and i.i.d. assumption of the error term can be nested in Assumption \ref{ass:sta}.
 
One crucial feature of Assumption \ref{ass:sta} is that it can accommodate endogenous covariates by allowing for arbitrary dependence structures between the fixed effects $\alpha_{ij}$ and the covariates $X_{it}$. Endogeneity is important in demand estimation because the price of a product could potentially depend on the unobserved heterogeneity of the product, such as the quality of the product or consumers' taste for the product. \cite*{chesher2013} and \cite{berry2014identification} provide more detailed discussions about the importance of allowing endogeneity in demand estimation. 
%Allowing for endogeneity has crucial practical meanings: as shown in \cite{luarn2003}, consumers' unobserved heterogeneity such as taste or loyalty to a brand depends on the product's characteristics such as prices, designs, or services. 

Of course, Assumption \ref{ass:sta} does impose some restrictions. For example, it excludes some dependence structures between $\epsilon_{it}$ and the covariate $X_{it}$. Consider that if $\epsilon_{it}$ only depends on $X_{it}$ for any period $t$, then $\epsilon_{is}$ may have a different distribution than $\epsilon_{it}$ when $X_{is}$ and $X_{it}$ take different values. However, some dependence structures between $\epsilon_{it}$ and $X_{it}$ are still allowed in Assumption \ref{ass:sta}: for example, if $\epsilon_{it}$ depends on covariates in a time-invariant form such as $\frac{1}{T}\sum_{t=1}^T X_{it}'\beta_0$,  Assumption \ref{ass:sta} still holds.

\subsection{Substitution Patterns}\label{subsec:sub}
This section discusses the relationship between two different definitions of substitution patterns. My paper uses the sign of $\Gamma(z)$ to represent the substitution relationship between two goods, which captures the incremental utility from consuming the bundle compared to consuming a single good. As shown in Lemma \ref{lem1}, this turns out to be equivalent to an alternative definition that was previously used in the literature.

%Now I introduce an alternative definition of substitution patterns that is widely used in the literature such as \cite{gentzkow2007}. In Lemma \ref{lem1}, an equivalence result between the two different definitions is established. 

The alternative definition of substitution patterns centers on how the demand for good $A$ (or $B$) is affected by an increase in the price of good $B$ (or $A$). The two goods are substitutes if the demand for good $A$ increases, complements if it decreases, and independent if the demand does not change. 
  Let $p_{jt}$ denote the price of good $j$ whose coefficient is nonzero, and let $\tilde{X}_{it}=X_{it} \setminus \{p_{Bt}\} $ denote the remaining covariates in $X_{it}$ excluding the price of good $B$.
I fix all other covariates $\tilde{X}_{is}=\tilde{X}_{it}=\tilde{x}$ over time and compare the conditional demand for good $A$ under different  prices $p_{Bs}\neq p_{Bt}$ of good $B$.  The demand for good $A$ comes from two sources: individuals who purchase only good $A$ and those who purchase the bundle $AB$. Let $D_{\ell}=\{\ell, AB\}$ collect all choices containing good $\ell \in\{A, B\}$. Let $\text{sign}(x)=\mathbbm{1}\{x>0\}-\mathbbm{1}\{x<0\}$ denote the sign function.

The substitution pattern $s_{AB}(z)$ conditional on the covariate $Z_i=z$ is defined as
\begin{equation*}
s_{AB}(z)\equiv \sign\left\{ \frac{ \Pr(Y_{is}\in D_A\mid p_{Bs}, p_{Bt}, \tilde{x}, z)-\Pr(Y_{it}\in D_{A}\mid p_{Bs}, p_{Bt}, \tilde{x}, z) }{p_{Bs}-p_{Bt}} \right\}.
\end{equation*}

The value of $s_{AB}(z) \in\{-1,0,1\}$ represents the complementarity relationship between goods $A$ and $B$. For consumers with the covariate $Z_i=z$, the two goods are substitutes if $s_{AB}(z) =1$, independent if $s_{AB}(z) =0$, and complements if $s_{AB}(z) =-1$. Under the aforementioned Assumptions \ref{ass:par}-\ref{ass:sta}, the value of $s_{AB}(z)$ is the same defined by any two periods $s\neq t$, and it is independent of other variables except $z$ since the complementarity term $\Gamma(z)$ depends only on $z$; therefore, $s_{AB}(z)$ is written as a function of only $z$.

%From the definition of $s_{AB}(Z_i)$, it can be directly identified from observed data by using variation in prices and fixing all other covariates.. Also, the value of $s_{AB}(z)$ only depends on the covariate $z$ but
It is often difficult to study substitution patterns directly from the definition of $s_{AB}(z)$. The term $s_{AB}(z)$ uses only variation in prices and requires the fixing of all other covariates. This may not be feasible since the other covariates may change simultaneously with prices or the covariates may include time-varying variables such as time dummies. 
In addition, variation in prices may not be available in some scenarios in which the prices of products are constant over time.  Moreover, as the definition $s_{AB}(z)$ involves conditional choice probabilities, directly estimating $s_{AB}(z)$ may perform poorly, especially when the dimension of covariates is large. 

The next lemma establishes the relationship between $s_{AB}(z)$ and the incremental utility $\Gamma(z)$.
\begin{lemma}\label{lem1}
Under Assumptions \ref{ass:par}-\ref{ass:sta}, the following holds for any $z$:
\begin{equation*}
\Gamma(z)s_{AB}(z)\leq 0.
\end{equation*}
\end{lemma}

Lemma \ref{lem1} shows that $s_{AB}(z)$ always has the opposite sign of the incremental utility term $\Gamma(z)$. This lemma implies that the sign of $s_{AB}(z)$ can be learned if the sign of the incremental utility $\Gamma(z)$ is identified. Therefore, identifying the complementarity term $\Gamma(z)$ is sufficient for studying substitution patterns defined by $s_{AB}(z)$. 
 
To illustrate the intuition of Lemma \ref{lem1}, I will focus on the case in which the incremental utility is positive, i.e., $\Gamma(z)> 0$. In this case, consumers with a small utility from a single good will still purchase the bundle since they can obtain additional positive utility from consuming the two goods together. When the price of good $B$ increases such that the utility of the bundle decreases, some consumers will switch from buying the bundle to buying the outside option since their utility from a single good is small. Therefore, the demand for good $A$ decreases, which implies $s_{AB}(z)\leq 0$.

A similar result to Lemma \ref{lem1} is shown in \cite{gentzkow2007} with cross-sectional data. The difference is that \cite{gentzkow2007} requires an independence condition between unobserved error terms and observed covariates, so his results do not apply to the case with endogenous covariates. My paper leverages the stationarity assumption, which allows for endogenous covariates. Therefore, Lemma \ref{lem1} shows that even with endogenous covariates, the relationship between the two definitions of substitution patterns ($\Gamma(z)s_{AB}(z)\leq 0$) still holds by exploiting intertemporal variation in covariates.

\section{Testing Complementarity}\label{subsec:test}

Before introducing the identification results, this section develops a method to test the complementarity relationship. This approach does not involve the estimation of any model parameters and directly constructs testable implications that only depend on observed variables.

I will focus on testing the complementarity between the two goods; the analysis for testing the substitutability is similar, which can be found in Appendix \ref{sec:test}.  For consumers with covariate $Z_i=z$, testing the complementarity relationship between goods is equivalent to testing $\Gamma(z)\geq 0$. The null hypothesis $H_0$ and alternative hypothesis $H_1$ are given as
\begin{equation*}
H_0: \Gamma(z)\geq 0 \qquad H_1: \Gamma(z)<0.
\end{equation*}

The main idea of testing the complementarity is that changing the covariate index $X_{iAt}'\beta_0$ of good $A$ will affect the demand for good $B$ in different directions, depending on their substitution/complementarity relationships. If the two goods are complements (under $H_0$), an increase in the covariate index of good $A$ will encourage consumers to purchase the bundle and thus increase the demand for good $B$. However, if the two goods are substitutes (under $H_1$), people will switch to choosing only good $A$ such that the demand for good $B$ decreases. Therefore, we can reject $H_0$ when a decrease in the demand for good $B$ is observed in this scenario.

However, the sign of the covariate index $X_{ijt}'\beta_0$ is unknown since it involves the unknown parameter $\beta_0$. Therefore, the first step of testing the complementarity is to learn the sign of covariate indices of two goods using observed variables $(X_{it}, Z_i, Y_{it})$. 

Let $P_{t}(\{j\} \mid x_s, x_t, z)$ denote the probability of choosing $j\in \mathcal{C}$ at time $t$ conditional on covariates $(X_{is}, X_{it})=(x_s, x_t)$ and $Z_i=z$, given as
\begin{equation*} 
\begin{aligned}
 P_{t}(\{j\} \mid x_s, x_t, z)=\Pr(Y_{it}=j\mid x_s, x_t, z).
 \end{aligned}
\end{equation*}
%\Pr\big(u_{ijt}\geq u_{ikt} \ \ \forall k\in \mathcal{C}  \mid x_s, x_t, z\big)

The conditional choice probability depends only on observed variables, which can be identified from data. Since the conditional distribution of error terms is the same over time (as per Assumption \ref{ass:sta}), any variation in choice probabilities can only arise from changes in covariate indices. From the variation in observed choice probabilities, we can infer the signs of the covariate indices and construct testable implications.

Let $\xi_{s, t}^{1}(x_{s}, x_{t}\mid z)$ denote an indicator for increasing probabilities of all choices $j\in \{A, B, AB\}$ conditional on $(X_{is}, X_{it})=(x_s, x_t)$ and given $Z_i=z$:
\begin{equation*}
\begin{aligned}
\xi_{s, t}^{1}(x_{s}, x_{t}\mid z) &= \mathbbm{1}\Big\{P_s(\{j\} \mid x_{s}, x_{t}, z)- P_t(\{j\} \mid x_{s}, x_{t}, z)\geq 0, \  \forall j \in \{A, B, AB\}   \Big\}. \\
%\xi_{s, t}^{2}(X_{is}, X_{it}) &= \mathbbm{1}\big\{P_s(j |X_{is}, X_{it})- P_t(j |X_{is}, X_{it})\geq 0,  \ \forall j \in \{A, AB, 0\}  \big\}  \\
\end{aligned}
\end{equation*}
%As shown in Section \ref{sec:testable}, when $\xi_{s, t}^{1}(x_s, x_t)=1$ which denotes increasing conditional probability of all choices $j\in\{A, B, AB\}$,  the covariate indices for the two goods should both increase:

When an increase in conditional probabilities of all choices $j\in \{A, B, AB\}$ is observed, it implies that the covariate indices of both goods increase: 
\begin{equation*}
\xi_{s, t}^{1}(x_s, x_t\mid z)=1 \Longrightarrow  x_{As}'\beta_0-x_{At}'\beta_0 \geq 0, \ x_{Bs}'\beta_0-x_{Bt}'\beta_0 \geq 0.
\end{equation*}

The above relationship exploits variation in conditional probabilities of multiple choices to identify the signs of covariate indices of both goods. If we observe an increase in the probability of only one choice, such as good A, it could be due to either an improvement in good A or a decline in good B, and we are unable to distinguish between the two scenarios. In contrast, if we observe an increase in the probabilities of all choices $\{A, B, AB\}$, it can only occur when both goods improve, enabling us to determine the signs of the variation in the covariate indices for both goods.

Now we are ready to establish testable conditions for the null hypothesis. With the null hypothesis of the two goods being complements, an increase in the covariate indices of both goods would result in an increase in demand for the two goods. This relationship generates testable implications for the null hypothesis.  If a decrease in demand for either good is observed, it suggests that the two goods are substitutes and the null hypothesis is rejected.

Recall that $D_{\ell}=\{\ell, AB\}$ for $\ell \in \{A, B\}$, and the conditional demand of good $\ell$ is expressed as 
\[\Pr(Y_{it}\in D_{\ell}\mid x_{s}, x_{t}, z)=E[\mathbbm{1}\{Y_{it}\in D_{\ell}\}\mid x_{s}, x_{t}, z].\]

%The following proposition provides testable implications for the complementarity.

\begin{prop}\label{prop:test1}
Under Assumptions \ref{ass:par}-\ref{ass:sta}, the following conditional moment inequalities hold under the null hypothesis $H_0$ given $Z_i=z$,
\begin{equation*}
E\Big[ \xi_{s, t}^{1}(x_{s}, x_{t}\mid z) \big(\mathbbm{1}\{Y_{is}\in D_{\ell}\}- \mathbbm{1}\{Y_{it}\in D_{\ell} \} \big)  \bigm| x_{s}, x_{t}, z\Big] \geq 0,
\end{equation*}
 for any $(x_s, x_t), \ell \in \{A, B\},  s\neq t \leq T$.
 
\end{prop}

Proposition \ref{prop:test1} provides testable implications for the null hypothesis $H_0$ by characterizing conditional moment restrictions that depend only on observed variables. Therefore, the null hypothesis can be tested by directly testing the above conditional moment inequalities. The conditions in the proposition apply to any two periods and any pair of covariates, allowing us to test for complementarity by exploiting the variation in covariates over time.

%The conditions in Proposition \ref{prop:test1} hold for any pair of covariates at any two periods so that variation in covariates over time has been exploited to test complementarity. 

%The analysis of testing substitutability is provided in Appendix \ref{sec:test}.

%This section provides a method to test the sign of the complementarity $\Gamma(z)$ without estimating any model parameters. However, we may be also interested in the value of the complementarity $\Gamma(z)$ as well as the utility coefficient $\beta_0$. Section \ref{sec:inde} characterizes sharp identification results for the parameters $(\beta_0, \Gamma)$.

%, which includes the utility coefficient $\beta_0$ and the complementarity function $\Gamma$
\section{Identification}\label{sec:inde}

Besides testing the sign of $\Gamma(z)$, we may also be interested in the value of the complementarity $\Gamma(z)$ as well as the utility coefficient $\beta_0$. 
This section establishes sharp identification results for the parameter $\theta_0=(\beta_0, \Gamma)$. The observed variables include the covariates $(X_{it}, Z_i)$ and consumers' choices $Y_{it}\in \mathcal{C}$ in each period. Since only the relative utility between choices matters for consumers' decisions, the parameter can be only identified up to a constant. Therefore, I normalize the first element $\theta^1$ of the parameter $\theta$ to be one for the following analysis: $\Theta=\{\theta: |\theta^{1}|=1\}$.

%For any subset $K\subset \mathcal{C}$, let $P_{t}(K\mid x_s, x_t, z)$ denote the conditional choice probability of  for $K\subset \mathcal{C}$

% For any subset $K\subset \mathcal{C}$, let $P_{t}(K\mid x_s, x_t, z)$ denote the conditional choice probability (CCP), which specifies the probability of $Y_{it}\in K$ at time $t$ conditional on covariates $(X_{is}, X_{it})=(x_s, x_t)$ and $Z_i=z$. This can be written as the probability of one choice in the set $K$ generating the highest utility among all choices:
% \begin{equation*} 
% \begin{aligned}
%  P_{t}(K\mid x_s, x_t, z) &\equiv\Pr(Y_{it}\in K \mid x_s, x_t, z)\\
%  &=\Pr\big(\exists j\in K \ \text{s.t.} \   \forall k\in \mathcal{C} \ u_{ijt}\geq u_{ikt}   \mid x_s, x_t, z\big).
%  \end{aligned}
% \end{equation*}

%Let $K\subseteq \mathcal{C}$ be any subset of the choice set. The conditional probability of $Y_{it}\in K$ at time $t$, given covariates $(X_{is}, X_{it})=(x_s, x_t)$ and $Z_i=z$, is denoted by $P_{t}(K\mid x_s, x_t, z)$. This probability represents the likelihood of a choice from the set $K$ generating the highest utility among all choices.

For any subset $K\subset \mathcal{C}$, let $P_{t}(K\mid x_s, x_t, z)$ denote the conditional probability of $Y_{it}\in K$ at time $t$ given covariates $(X_{is}, X_{it})=(x_s, x_t)$ and $Z_i=z$; that is,  the probability of existing one choice in the set $K$ generating the highest utility among all choices:
\begin{equation*} 
\begin{aligned}
 P_{t}(K\mid x_s, x_t, z) &\equiv\Pr(Y_{it}\in K \mid x_s, x_t, z)\\
 &=\Pr\big(\exists j\in K \ \text{s.t.} \   \forall k\in \mathcal{C} \ u_{ijt}\geq u_{ikt}   \mid x_s, x_t, z\big).
 \end{aligned}
\end{equation*}
When $K=\{j\}$ is a singleton, this reduces to the conditional choice probability (CCP) of $j$. The main idea of the identification analysis is to derive identifying restrictions of the true parameter $\theta_0$ from intertemporal variation in conditional choice probabilities across two different periods. All parameters satisfying those identifying restrictions form an identified set for the true parameter.

Let $\delta_{\ell t}=x_{\ell t}'\beta_0$ denote the covariate index for good $\ell \in \{A, B\}$ given $X_{i\ell t}=x_{\ell t}$. Let $\delta_{ABt}=\delta_{At}+\delta_{Bt} $ and $\delta_{Ot}=0$ denote the covariate indices for bundle $AB$ and the outside option, respectively. Let $\Delta_{s,t} \delta_{j}=\delta_{js}-\delta_{jt}$ denote the change in the covariate index for choice $j\in \mathcal{C}$ between periods $s$ and $t$.

In models assuming that consumers can buy only one good at a time, two goods can only be substitutes. Since the complementarity relationship is known, the only unknown factor affecting conditional choice probabilities is variation in covariate indices of all choices. My paper allows for the possibility that two goods can be either substitutes ($\Gamma(z)<0$) or complements ($\Gamma(z)>0$) and the complementarity relationship is unknown. Therefore, two unknown sources are affecting conditional choice probabilities in this paper: one is changes in covariate indices and the other is the complementarity relationship between the two goods. Distinguishing between the two sources poses a challenge for the identification analysis.
% which sets it apart from existing literature.
 
The following proposition characterizes the identifying restrictions for the parameter $\theta_0$ under Assumptions \ref{ass:par}-\ref{ass:sta}. Let $C_1 \lor C_2$ mean that either condition $C_1$ or $C_2$ holds or both hold, and let $C_1 \land C_2$ mean that both $C_1$ and $C_2$ hold.

\begin{prop} \label{prop1}
Under Assumptions \ref{ass:par}-\ref{ass:sta}, the following conditions hold for any $(x_s, x_t, z)$ and any $s\neq t$: 

\begin{enumerate}

\item[{(1)}] comparisons of CCP of choice $j\in \mathcal{C} $: 
\begin{equation}\tag{ID1} \label{id1}
\begin{aligned}
P_s(\{j\}\mid x_s, x_t, z)>  P_t(\{j\} \mid x_s, x_t, z)  \Longrightarrow   \exists k\neq j \ \text{s.t.} \  \Delta_{s,t} \delta_{j}> \Delta_{s,t} \delta_{k};
\end{aligned}
\end{equation}

\item[{(2)}] comparisons of the demand for good $\ell \in \{A, B\}$:% and let $\ell_{-1}\neq \ell \in \{A, B\}$, 
\begin{equation}\tag{ID2} \label{id2}
\begin{aligned}
P_s\big( &\{\ell, AB\}\mid x_s, x_t,  z \big)>P_t \big(\{\ell, AB\}\mid x_s, x_t, z\big) \Longrightarrow \\
&\{ \Delta_{s,t}\delta_{\ell} > 0\} \lor \Big \{ \Delta_{s,t} (\delta_{\ell}+ \sign(\Gamma(z)) \delta_{\ell_{-1} })> 0,  |\Gamma(z)|> -\Delta_{s,t}\delta_{\ell} \Big\},
\end{aligned}
\end{equation}
where $\ell_{-1}\in\{A, B\}$ and $\ell_{-1}\neq \ell$;

\item[{(3)}] comparisons of the sum of CCP of two choices:
\begin{equation} \tag{ID3} \label{id3}
\begin{aligned}
 P_s(\{AB\}\mid  x_s,  x_t, z)&+P_t(\{O\}\mid x_s, x_t, z)> 1 \Longrightarrow \\
 \Big\{ \Gamma(z)&>-\min\{\Delta_{s,t} \delta_{A}, \Delta_{s,t} \delta_{B} \}  \Big\} \land \{ \Delta_{s, t} ( \delta_{A}+ \delta_{B}) >0\},\\
P_s(\{A\}\mid x_s,  x_t, z)& +P_t(\{B\}\mid x_s, x_t, z)> 1 \Longrightarrow  \\ 
\Big\{\Gamma(z)&<\min\big\{\Delta_{s,t} \delta_{A}, -\Delta_{s,t} \delta_{B}\} \Big\} \land \{  \Delta_{s, t}( \delta_{A}- \delta_{B}) > 0 \}. \\
 \end{aligned}
 \end{equation} 
 
\end{enumerate}
\end{prop}

Proposition \ref{prop1} characterizes identification restrictions for the parameter $\theta_0$ from comparisons of conditional choice probabilities across two periods that can be identified from data. The identifying restrictions for $\theta_0$ in Proposition \ref{prop1} are free from unobserved terms, such as the fixed effects $\alpha_i$ and the error term $\epsilon_{it}$. As the results hold for any fixed length $T$ of panel data, we can utilize variation in conditional choice probabilities for any two periods to identify $\theta_0$ and take intersections of the identified sets. Later I will formulate conditional moment inequalities based on the identifying restrictions in Proposition \ref{prop1}, which can be used to conduct estimation and inference for the parameter $\theta_0$.

%Since the above results hold for any fixed length $T$ of panel data, I can use variation in conditional choice probabilities for any two periods to identify $\theta_0$ and take intersections of the identified sets. Later I will formulate conditional moment inequalities based on the identifying restrictions in Proposition \ref{prop1}, which can be used to conduct estimation and inference for the parameter $\theta_0$. 

Condition \eqref{id1} in Proposition \ref{prop1} contains the identifying restrictions for the coefficient $\beta_0$. The intuition of this result is described as follows: if the conditional probability of selecting choice $j$ increases, then it is impossible that choice $j$ becomes worse (in terms of the covariate index) compared to all other choices. Therefore, it can be inferred that the covariate index for choice $j$ should increase relative to at least one other choice.
%This identification result is similar to the results in \cite{pakes2019}, but the difference is that the bundle $AB$ is also a feasible choice in this paper.  

The remaining two conditions in Proposition \ref{prop1} provide novel identification results for the complementarity parameter $\Gamma(z)$. Condition \eqref{id2} identifies the sign of the complementarity $\Gamma(z)$ and bounds its absolute value by comparing the conditional demand of the two goods over time. Condition \eqref{id3} establishes both lower and upper bounds for the complementarity $\Gamma(z)$ using the sum of probabilities of two different choices over two periods. Next,  I will explain the intuition behind the two conditions. 

Condition \eqref{id2} mainly exploits the idea that the relationship between the covariate index of one good and the demand for the other good is different under different complementarity relationships between the two goods. When the two goods are complements, an increase in the covariate index of good $A$ will incentivize consumers to buy the bundle $AB$, resulting in an increase in demand for good $B$. So if a decrease in demand for either good is observed, the two goods must be substitutes $(\Gamma(z)< 0)$. Similarly, when the covariate index for good $A$ decreases and the covariate index for good $B$ increases along with observing an increase in the demand for good $A$, the two goods are identified as complements $(\Gamma(z)>0)$.

Condition \eqref{id3} in Proposition \ref{prop1} can bound the value of the complementarity $\Gamma(z)$ by looking at the sum of conditional probabilities of two choices over time. For example, when the sum of the conditional probabilities of buying two goods together and neither good is large, then a lower bound for the complementarity $\Gamma(z)$ is established. The intuition is that when good $A$ becomes less attractive, consumers will switch from buying two goods together to buying neither good if the two goods are complements, but they will switch to buying only good $B$ if the two goods are substitutes. Therefore, if a large probability of buying two goods together and neither good is observed, we can infer the complementarity between the two goods and provide a lower bound for $\Gamma(z)$. Similarly, if the sum of the conditional probabilities of buying a single good is large, then the upper bound for the value of $\Gamma(z)$ can be obtained.

% The intuition is that if the sum of conditional probabilities of buying two goods together and neither good is large, then the complementarity $\Gamma(z)$ cannot be too small because otherwise consumers will prefer to buy a single good instead of buying the two goods together. Therefore, the lower bound for the value of the complementarity can be established in this case. 

The identifying restrictions \eqref{id1}-\eqref{id3} in Proposition \ref{prop1} characterize an identified set $\Theta_I$ for $\theta_0$, which is defined as
\begin{equation*}
\Theta_I=\{ \theta \in \Theta: \text{conditions} \  \eqref{id1}-\eqref{id3} \  \text{hold with} \ \theta \ \text{in place of} \ \theta_0 \}. 
\end{equation*}

%The next theorem will show the sharpness of the identified set $\Theta_0$.
\begin{thm}\label{thm1}
Under Assumptions \ref{ass:par}-\ref{ass:sta}, the identified set $\Theta_I$ is sharp.
\end{thm}

Theorem \ref{thm1} shows that conditions \eqref{id1}-\eqref{id3} have exhausted all available information from the observed data for the parameter $\theta_0$. 
%The identified set $\Theta_I$ coincides with the identified set for $\beta_0$ in \cite{pakes2019} when assuming no bundles $\Gamma(z)=-\infty$ for any $z$ and there are only two goods.  
% However, the results in \cite{pakes2019} are no longer sharp if we allow for any possible value of $\Gamma(z)$ since Assumption \ref{ass:par} imposes an additional structure on the incremental utility. This structure allows us to identify the sign and bound the value of the complementarity.  Theorem \ref{thm1} characterizes the sharp identified set for both the complementarity $\Gamma(z)$ and the coefficient $\beta_0$.
% It is sufficient to exploit a pairwise comparison of conditional choice probabilities since Assumption \ref{ass:sta} only imposes a pairwise stationarity condition on the error term.
The proof of the sharpness is conducted through direct construction. For any parameter in the identified set $\Theta_I$, I construct an underlying DGP that satisfies Assumptions \ref{ass:par}-\ref{ass:sta} and matches the observed conditional choice probabilities, which shows the sharpness of the identified set $\Theta_I$. 
 The main challenge of the construction is that the unknown DGP involves conditional distributions over the whole space of unobserved error terms, which are infinite dimensional.

I address this difficulty by constructing ``choice sets," which are collections of unobserved terms such that a single choice is selected conditional on covariates. It is sufficient to focus on constructing the distributions on the choice sets because their distributions determine the observed choice probabilities. The number of choice sets is finite due to the finite number of choices; accordingly, I  only need to assign probabilities on the finite number of sets, which simplifies the construction. 
Then the paper shows that for any parameter in the identified set $\Theta_I$, there exists a conditional distribution on the choice sets that satisfies the assumptions and generates the observed choice probabilities. The construction of the probabilities on the choice sets depends on the sign of the complementarity $\Gamma(z)$ as well as the covariate index $\Delta_{s, t} \delta_j$, which is discussed in detail in Appendix \ref{proof:thm1}.

Some discussion of Theorem \ref{thm1} is in order.  First, similar to \cite{pakes2019}, the identification analysis in Proposition \ref{prop1} and the sharpness result for $\Theta_I$ can be extended to a more general model,  $u_{i\ell t}=f(X_{i\ell t}, \beta_0)+g(\alpha_{i\ell}, \epsilon_{i\ell t})$, where $f$ is a known function up to a finite-dimensional parameter $\beta_0$ and $g$ is a function that can be unknown to econometrician. This utility function allows for infinite dimensional fixed effects and idiosyncratic shocks as well as admits arbitrary interactions between them. Appendix \ref{subsec:nonsep} also discusses a class of utility functions that can be nonseparable between covariates and unobserved fixed effects/error terms. Second, the identified set $\Theta_I$ employs only marginal choice probabilities at each period yet it is shown to be sharp. Therefore, joint choice probabilities over different periods do not provide any additional information for the parameter $\theta_0$. Moreover, the sharpness result exploits the correlations of error terms across choices and over time. So if one is willing to impose additional assumptions on the dependence structure (e.g., i.i.d.), then the identified set $\Theta_I$ could be further tightened.

\subsection{Point Identification}\label{point}

This section studies the conditions under which the model parameters can be point identified up to scale. The analysis depends on the specification of the additional utility term $\Gamma(Z_i)$. I focus on point identification under a linear specification of the complementarity: $\Gamma(Z_i)=Z_i'\gamma_0$.

For simplicity of notation, I consider a two-period model ($T=2$) to illustrate the idea.\footnote{With more than two periods, it is straightforward that point identification can be achieved when there exists a pair of periods that satisfy Assumptions \ref{ass:ep}-\ref{ass:z}.} Let $\Delta X_{i\ell }= X_{i\ell 2}-X_{i\ell 1}$ denote the change in observed covariates for consumer $i$ and good $\ell\in \{A, B\}$ over the two periods, and let $\Delta X_i=(\Delta X_{iA}, \Delta X_{iB})$ collect the changes in covariates for the two goods.  I use a superscript $k$ to denote the $k$th element of a vector, e.g., $\Delta X_{iA}^{k}$ represents the $k$th element of the vector $\Delta X_{iA}$. 

I first introduce sufficient conditions for point identification of the coefficient $\beta_0$. The coefficient $\beta_0$ can only be point identified up to scale since multiplying consumers' utilities of all choices by a positive constant will not change observed choices. 

\begin{ass}\label{ass:ep}
%The support of the conditional density of $\epsilon_{it}$ given $(X_{i1}, X_{i2}, Z_i, \alpha_i)$ is $\mathbb{R}^2$.
The density of $\epsilon_{it}$ conditional on $(X_{i1}, X_{i2}, Z_i, \alpha_i)$ is positive everywhere on $\mathbb{R}^2$ for $t\in \{1, 2\}$.

%Conditional on $(X_{i1}, X_{i2}, Z_i, \alpha_i)$, $\epsilon_{it}$ is continuously distributed on $\mathbb{R}^2$ for $t\in \{1, 2\}$.
\end{ass}

\begin{ass}\label{ass:x}
 For any $\ell\in \{A, B\}$, there exists $k_{\ell}$ that satisfies $\beta_0^{k_{\ell}}\neq 0$. Let $\Delta \tilde{X}_{i}=\Delta X_i\setminus (\Delta X_{iA}^{k_A}, \Delta X_{iB}^{k_B})$ denote the remaining elements in $\Delta X_{i}$. The density of $(\Delta X_{iA}^{k_A}, \Delta X_{iB}^{k_B})$ conditional on $(\Delta \tilde{X}_{i}, Z_i)$ is positive everywhere on $\mathbb{R}^2$. Furthermore, the support of $\Delta X_{i\ell}$ is not contained in any proper linear subspace of $\mathbb{R}^{d_x}$.

 %For any $\ell\in \{A, B\}$, there exists $k_{\ell}$ that satisfies $\beta_0^{k_{\ell}}\neq 0$. Let $\Delta \tilde{X}_{i}=\Delta X_i\setminus (\Delta X_{iA}^{k_A}, \Delta X_{iB}^{k_B})$ denote the remaining elements in $\Delta X_{i}$. The conditional density of $(\Delta X_{iA}^{k_A}, \Delta X_{iB}^{k_B})$ given $(\Delta \tilde{X}_{i}, Z_i)$ is positive everywhere on $\mathbb{R}^2$. Furthermore, the support of $\Delta X_{i\ell}$ is not contained in any proper linear subspace of $\mathbb{R}^{d_x}$.
 
\end{ass}
 %assumes that the conditional density of $\epsilon_{it}$ is positive everywhere on $\mathbb{R}^2$.  This 
 Assumption \ref{ass:ep} eliminates the uninformative scenario where conditional choice probabilities remain unchanged despite changes in the covariate indices over time.
Assumption \ref{ass:x} is a support condition on the covariate $\Delta X_i$. It requires at least one covariate for each good to have large support, while the support of the remaining covariates is unrestricted. The large support condition guarantees that 
 there is sufficient variation in the covariate over time such that the true parameter can be distinguished from any other candidate parameters.
 
Under these assumptions, $\beta_0$ can be point identified by using the first identifying restriction (condition \eqref{id1}) in Proposition \ref{prop1}. For any parameter $b$ such that $b\neq k\beta_0$, $\forall k>0$, the large support condition in Assumption \ref{ass:x} implies that there exists one value $\Delta x_{\ell}$ of the covariate such that the covariate index $\Delta x_{\ell}' \beta$ has different signs under the true parameter $\beta_0$ and the candidate parameter $b$. The conditional choice probabilities then change in different directions under $\beta_0$ and $b$ so that the parameter $\beta_0$ is identified. 
For example, suppose that the covariate index satisfies $\Delta  x_{\ell}' \beta_0 >0$ and $\Delta  x_{\ell}'  b <0$ for any $\ell \in \{A, B\}$. Then under Assumption \ref{ass:ep}, the conditional choice probability of buying bundle $AB$ will strictly increase under the true parameter $\beta_0$, but strictly decrease under the parameter $b$. Therefore, $\beta_0$ is identified. 

Since $\beta_0$ is point identified, the sign of the covariate index $\Delta X_{ij}'\beta_0$ is also identified. Next, I present the conditions for point identification (up to scale) of the complementarity parameter $\gamma_0$.

\begin{ass}\label{ass:z}
There exists $k$ such that $\gamma_0^{k}\neq 0$. Let $\tilde{Z}_i=Z_i\setminus Z_i^k$ denote the remaining elements in $Z_i$. The density of $Z_i^{k}$ conditional on $(X_{i1}, X_{i2},\tilde{Z}_i )$ is positive everywhere on $\mathbb{R}$. Furthermore, the support of $Z_i$ is not contained in any proper linear subspace of $\mathbb{R}^{d_z}$.

\end{ass}

Similar to Assumption \ref{ass:x}, this assumption requires a large support restriction on the covariate $Z_i$. Based on condition \eqref{id2} in Proposition \ref{prop1}, the sign of the complementarity $Z_i'\gamma_0$ is identified from intertemporal variation in the conditional demand for the two goods. Since the sign of $Z_i'\gamma_0$ is identified, the parameter $\gamma_0$ is also point identified up to scale under the large support assumption. The analysis for $\gamma_0$ is similar to the coefficient $\beta_0$. For any candidate parameter $\tilde{\gamma}$ such that $\tilde{\gamma}\neq k\gamma_0$, $\forall k>0$, Assumption \ref{ass:z} ensures that there exists some value of the covariate $Z_i$ such that the sign of the complementarity $Z_i'\gamma$ is different under the true parameter $\gamma_0$ and the candidate parameter $\tilde{\gamma}$. Thus, the parameter $\gamma_0$ can be point identified.

\begin{thm}\label{thm2}
Under Assumptions \ref{ass:par}-\ref{ass:z} and $\Gamma(Z_i)=Z_i'\gamma_0$, the parameters $\beta_0$ and $\gamma_0$ are point identified up to scale.
\end{thm}

%Theorem \ref{thm2} establishes the point identification results under the large support assumptions of covariates and the linear specification of the complementarity. For instance, if a covariate is the price of a product, the large support assumption requires that there is large variation in prices over time. This is arguably a strong assumption, but unfortunately it is necessary for point identification 

%\cite{manski1987} shows that point identification fails in binary choice models with bounded support of covariates. 

%Without the large support assumption, I characterize the sharp identified set $\Theta_I$ for $\theta_0$ in Theorem \ref{thm1} without restricting the support of covariates.

\section{Conditional Moment Inequalities}\label{sec:esti} 

The identified set $\Theta_I$ characterized by conditions \eqref{id1}-\eqref{id3} in Proposition \ref{prop1} is abstract and it is a challenging task to check whether every candidate parameter satisfies all of the identifying conditions. This section develops an alternative characterization of the identified set $\Theta_I$ by constructing conditional moment inequalities of the parameter.  Based on this characterization, the literature has developed many methods to do inference for conditional moment inequalities (e.g., \cite{andrews2013}, \cite*{chernozhukov2013}, and \cite*{armstrong2015}).

The identification conditions \eqref{id1}-\eqref{id3} in Proposition~\ref{prop1} share a similar structure, which involves deriving restrictions for the parameter $\theta_0$ through intertemporal comparisons of conditional choice probabilities. I focus on the first condition \eqref{id1} in Proposition \ref{prop1} to describe the idea of constructing conditional moment inequalities. 
Let $W_{ist}=(X_{is}, X_{it}, Z_i)$ collect all of the covariates at the two periods $(s, t)$, and let $w_{st}=(x_{s}, x_t, z)$ denote one realization of the covariate $W_{ist}$.

Condition \eqref{id1} exploits comparisons of the conditional probability of a single choice $j\in \mathcal{C}$ to derive restrictions for the parameter. Let $\lambda_{s, t}^{j}(w_{st}, \theta)$ denote the indicator index of the identifying restriction in condition \eqref{id1}, defined as
\begin{equation*}
\lambda_{s,t}^{j}(w_{st}, \theta) =
\mathbbm{1} \big\{ \exists \ k \neq j \  \text{s.t.} \  \Delta_{s, t} x_{j}'\beta > \Delta_{s, t} x_{k}'\beta \}. \\
\end{equation*}

Condition \eqref{id1} derives the identifying restriction $\lambda_{s,t}^{j}$ from a positive variation in the conditional probability of selecting choice $j$ over time:
\begin{equation*}
P_s(\{j\}\mid w_{st})-P_t(\{j\}\mid w_{st})> 0 \Longrightarrow \lambda^{j}_{s,t}(w_{st}, \theta_0)=1.
\end{equation*}

The contraposition of the above condition is presented as follows: if the identifying restriction $\lambda^{j}_{s,t}$ does not hold, then the variation in the conditional probability of selecting choice $j$ is nonpositive.
\begin{equation*}
\lambda^{j}_{s,t}(w_{st}, \theta_0)=0 \Longrightarrow  P_s(\{j\}\mid w_{st})-P_t(\{j\}\mid w_{st})\leq 0.
\end{equation*}

Plugging into the definition of the conditional choice probability $P_t(\{j\}\mid w_{st})=E[\mathbbm{1}\{Y_{it}=j \} \mid W_{ist}=w_{st}]$, the above condition leads to the following conditional moment inequality for any $w_{st}$,
\begin{equation*}\label{momine1}
g_{s,t}^{j}(w_{st}, \theta_0)=E\big[ (1-\lambda^{j}_{s,t}(w_{st}, \theta_0))(\mathbbm{1} \{Y_{is}=j \}-\mathbbm{1}\{Y_{it}=j  \} ) \mid W_{ist}=w_{st}\big]\leq 0.
\end{equation*}

The above conditional moment inequality holds since either the binary index holds $\lambda^{j}_{s,t}(w_{st}, \theta_0)=1$ so that the moment function $g_{s,t}^{j}$ is zero or the binary index does not hold $\lambda_{s,t}(w_{st}, \theta_0)=0$ implying that the function $g_{s,t}^{j}$ is nonpositive.  I provide an equivalent characterization to condition \eqref{id1} using conditional moment inequalities. The characterization for conditions \eqref{id2}-\eqref{id3}  can be constructed similarly.

%Next I define the binary indicator of identifying restrictions for the parameter in conditions \eqref{id2}-\eqref{id3}. 

Condition \eqref{id2} derives restrictions of the parameter from comparisons of the demand for good $\ell \in \{A, B\}$. The indicator $\lambda_{s,t}^{D_{\ell}}(w_{st}, \theta)$ of the identifying restriction in condition \eqref{id2} is defined as follows,  let $\ell_{-1} \in \{A, B\}$ and $\ell_{-1} \neq \ell$,
\begin{equation*}
\begin{aligned}
\lambda_{s,t}^{D_{\ell}}(w_{st}, \theta)=\mathbbm{1}  \Bigg\{ & \{\Delta_{s,t} x_{\ell}'\beta > 0 \}  \\
&  \lor  \Big\{ \Delta_{s,t} (x_{\ell} +\sign(\Gamma(z)) x_{\ell_{-1} })'\beta > 0, |\Gamma(z)| >- \Delta_{s,t}x_{\ell}'\beta  \Big \} \Bigg\}.
\end{aligned}
\end{equation*}

From comparisons of the demand for good $\ell\in \{A, B\}$, the conditional moment inequality can be constructed as follows:
\begin{equation*}
g^{D_{\ell}}_{s, t}(w_{st}, \theta_0)=E\big[(1-\lambda_{s,t}^{D_{\ell}}(w_{st}, \theta_0))(\mathbbm{1} \{Y_{is}\in D_{\ell} \}-\mathbbm{1}\{Y_{it}\in D_{\ell} \} ) \mid W_{ist}= w_{st}\big]\leq 0.
\end{equation*}

Condition \eqref{id3} derives lower and upper bounds for the complementarity $\Gamma(z)$ from the sum of conditional probabilities of two choices over two different periods. The binary indices of the identifying restrictions in condition \eqref{id3} are defined as
\begin{equation*}
\begin{aligned}
\lambda_{s,t}^{L}(w_{st}, \theta)&=\mathbbm{1} \Big\{ \big\{ \Gamma(z )> -\min \{\Delta_{s,t} x_{A}'\beta,\Delta_{s,t} x_{B}'\beta \} \big\} \land \{ \Delta_{s, t} ( x_{A}+ x_{B})'\beta > 0 \} \Big \},  \\
\lambda_{s,t}^{U}(w_{st}, \theta)&=\mathbbm{1} \Big\{\big\{ \Gamma(z)<\min\{\Delta_{s,t} x_{A}'\beta, -\Delta_{s,t} x_{B}'\beta\} \big\} \land \{\Delta_{s, t}( x_{A}- x_{B})'\beta > 0\} \Big\}.
\end{aligned}
\end{equation*}

Similarly, the conditional moment inequalities are constructed as follows based on condition \eqref{id3} in Proposition \ref{prop1}: 
\begin{equation*}\label{momine3}
\begin{aligned}
%&g^{D_{\ell}}_{s, t}(w_{st}, \theta_0)=E\big[(1-\lambda_{s,t}^{D_{\ell}}(w_{st}, \theta_0))(\mathbbm{1} \{Y_{is}\in D_{\ell} \}-\mathbbm{1}\{Y_{it}\in D_{\ell} \} ) \mid W_{ist}= w_{st}\big]\leq 0, \\
&g^{L}_{s, t}(w_{st}, \theta_0)=E\big[(1-\lambda_{s,t}^L(w_{st}, \theta_0))(\mathbbm{1} \{Y_{is}=AB \}+\mathbbm{1}\{Y_{it}=O \}-1 ) \mid W_{ist}=w_{st}\big]\leq 0,\\
&g^{U}_{s, t}(w_{st}, \theta_0)=E\big[(1-\lambda_{s,t}^U(w_{st}, \theta_0))(\mathbbm{1} \{Y_{is}=A \}+\mathbbm{1}\{Y_{it}=B \}-1 ) \mid W_{ist}=w_{st}\big]\leq 0.\\
\end{aligned}
\end{equation*}

I have developed conditional moment inequalities that are equivalent to the identifying conditions \eqref{id1}-\eqref{id3} in Proposition \ref{prop1}.
Let $g_{s, t}=(\{g_{s,t}^{j}\}_{j\in \mathcal{C}}, g_{s,t}^{D_{A}}, g_{s, t}^{D_B}, g_{s,t}^{L}, g_{s, t}^{U})'$ denote a vector of all conditional moment functions. The identified set $\Theta_I$ is characterized by the set of parameters satisfying the conditional moment inequalities.
\begin{prop}\label{prop2}
Under Assumptions \ref{ass:par}-\ref{ass:sta}, the following holds:
\begin{equation*}
\Theta_I=\{\theta\in \Theta: g_{s, t}(w_{st}, \theta)\leq 0 \quad \forall w_{st},\  \forall s, t\leq T \}.
\end{equation*}
\end{prop}

Proposition \ref{prop2} characterizes the identified set using conditional moment inequalities. For estimation and inference, we need to specific a parametric form for the function $\Gamma$ (such as a linear function) so that the dimension of the unknown parameter is finite. Then, the estimation and inference for the parameter can be conducted using methods in the literature developed for conditional moment inequalities. 
%Since only the relative utility between choices matters for consumers' decisions, the parameter can be only identified up to a constant. Therefore, I normalize the first element $\theta^1$ of the parameter $\theta$ to be one for the following analysis: $\Theta=\{\theta: \theta^{1}=1\}$.

\section{Simulation Study}\label{sec:simu}

This section will compare the method in this paper with a parametric method via Monte Carlo simulation. The parametric method imposes distributional assumptions over $\epsilon_{ijt}$ and a linear structure over $\alpha_{ij}$, which will be described in detail later. The simulation results demonstrate that misspecifications in either parametric distributions or dependence structures between covariates and fixed effects lead to misleading estimators for the complementarity parameter.  

%\subsection{Point Identification}

I study a linear specification of the complementarity $\Gamma(Z_i)=Z_i \gamma_0$, and look at a two period model $T=2$. Section \ref{point} has established point identification results under large support of covariates, so this section focuses on the case where the parameter is point identified. More simulation results are presented in Appendix \ref{sec:sim_more} regarding the performance of the point estimator under longer panels ($T\geq 2$) and the set estimator with bounded support of covariates.

 I implement the criterion function approach in \cite*{shi2018} for estimation. The criterion function can be developed as follows based on conditional moment inequalities in Proposition \ref{prop2}:
\begin{equation*}
\begin{aligned}
\Omega(\theta)=\sum_{s\neq t\leq T }E\Big[ \Vert \max\{ (g_{s,t}(W_{ist}, \theta), 0\} \Vert_1 \Big]
\geq \Omega(\theta_0)=0.
\end{aligned}
\end{equation*}
where $||x||_1=\sum_j |x^j|$ for a vector $x=(x^1,...,x^j)'$.

Similar to \cite*{shi2018}, a two-step estimator is developed based on the above criterion function. The first step estimates the conditional choice probability $P_{t}(\{j\} \mid w_{st})$ using a nonparametric estimator $\hat{P}_{t}(\{j\} \mid w_{st})$. I use a single layer artificial neural network estimator and the asymptotic property of this estimator has been established in \cite{chen1999}. The neural network estimator is computationally easy to implement and there is a readily used package for the estimator (\cite{bischl2016}). Let $\hat{g}_{s, t}$ denote the estimated moment function that replaces the conditional choice probability $P_{t}(\{j\} \mid w_{st})$ with its estimator $\hat{P}_{t}(\{j\} \mid w_{st})$, then the sample objective function $\hat{\Omega}_N(\theta)$ is constructed as follows:
%There could be a set of parameters which minimize the sample objective function, I just randomly pick one point from the set since any point estimator will converge to the true parameter. Also the set of minimizers is convex so that all points are close to each other and have similar performance. 

 \begin{equation*}
\begin{aligned}
\hat{\Omega}_N(\theta)=\frac{1}{N}\sum_i^{N} \sum_{ s\neq t\leq T}  \Vert\max\{ \hat{g}_{s,t}(W_{ist}, \theta), 0\} \Vert_1.
\end{aligned}
\end{equation*}

The second-step estimator for the parameter is obtained by minimizing the sample objective function $\hat{\Omega}_N$. Since the parameter $\beta_0$ and $\gamma_0$ can only be point identified up to a scale, the first element of the two parameters is normalized as one. This normalization is also used for other approaches to compare these methods fairly.
% In the simulation, I also assign a positive weight $G(x)=[2 \Phi(x)-1]\mathbbm{1}\{x\geq 0\}$ for intertemporal variation in choice probabilities, where $\Phi$ is the CDF of the standard normal distribution. Through simulations, it can be shown that this weight function can increase the performance of the estimator compared to a uniform weight.
%When point identification is achieved (considered in the simulation), $\hat{c}_N$ is set to be zero and the second-step estimator for the parameter is obtained by minimizing the sample objective function. When only partial identification is achieved, $c_N$ is chosen proportional to $ log(N)$.

To better evaluate the performance of the two-step estimator (Two-Step Est.) in this paper,  I implement a parametric estimator (Parametric Est.) using the method of simulated moments for comparison. For this parametric estimator, the error terms $\epsilon_{ijt}$ are assumed to follow a standard Gumbel distribution, independent across choices and periods, and also independent of all covariates. I allow the fixed effects $\alpha_{ij}$ to depend on covariates through a linear specification: $\alpha_{ij}=\eta_0+\bar{X}_{ij}'\eta_1+v_{ij}$, where $\bar{X}_{ij}=\frac{1}{T} \sum_{t} X_{ij t}$ denote the average covariates and $v_{ij}\sim \mathcal{N}(0, 1)$ follows standard normal distribution and independent of all covariates.
 This parametric estimator is $\sqrt{N}$ consistent when its assumptions are all correct, while could be inconsistent if either the parametric distribution or the linear model of the fixed effects is misspecified. 

For the coefficient $\beta_0$, I also evaluate the performance of two other estimators for comparison that do not allow for the purchase of bundles $\Gamma_{it}=-\infty$.
One estimator is Chamberlain's conditional fixed-effect logit estimator (FE Logit Est.). This estimator assumes $\epsilon_{ijt}$ to follow standard Gumbel distribution while leaving the distribution of the fixed effects $\alpha_{ij}$ unrestricted. 
The other estimator is the semiparametric estimator (Semi. Est.) which is developed under the stationarity assumption but assumes no bundles. Therefore, this estimator only uses conditional choice probabilities of $\{A, B, O\}$ to identify the coefficient $\beta_0$. 

Now I describe the simulation setup. Let $d_x=2$ and $d_z=2$ denote the dimension of the covariates $X_{it}$ and $Z_i$ respectively.  
In each simulation,  $X_{i\ell t}$ is drawn from the normal distribution $\mathcal{N}(0, d_x)$, independently across choices $\ell\in \{A, B\}$ and time $t\leq T$.  Let the first element of $Z_i$ be drawn from $\mathcal{N}(2, 2)$ and the second element from $\mathcal{N}(0, 1)$. 
The true parameters are set as: $\beta_0=\gamma_0=(1,1)$. 

I study four different designs of the error terms $\epsilon_{ijt}$ and fixed effects $\alpha_{ij}$. The first design considers the correct specification for the parametric estimator: $\epsilon_{ij}$ follows a Gumbel distribution and the fixed effects are specified as $\alpha_{ij}=\bar{X}_{ij}'\beta_0/2+ v_{ij}$.
In the second design, the error term $\epsilon_{it}$ follows a bivariate normal distribution with the correlation $\rho=-0.7$. So the parametric distribution of  $\epsilon_{it}$ is misspecified in this design. In the third design, I allow the fixed effects $\alpha_{ij}$ to depend on the covariates of the other good in a non-additive form:  $\alpha_{ij}=(\bar{X}_{ij}/2-\bar{X}_{ik})'\beta_0*(1+v_{ij})$ for $j\in \{A, B\}$ and $k\neq j \in \{A, B\}$. In this design, the parametric estimator assumes a wrong model for the fixed effects $\alpha_{ij}$.
The last design combines the second and third design, which considers both a misspecified distribution of $\epsilon_{it}$ and a misspecified model of $\alpha_{ij}$ for the parametric estimator. The following summarizes the four designs:
\begin{itemize}
\item Design 1: correct specification 
\begin{equation*}
\begin{aligned}
\epsilon_{ij t}&\sim \text{Gumbel}(0, 1), \\
\alpha_{ij}&=\bar{X}_{ij}'\beta_0/2+v_{ij}, \quad  \text{where} \ v_{ij}\sim \mathcal{N}(0, 1).
\end{aligned}
\end{equation*}

\item Design 2: misspecified distribution
\begin{equation*}
\begin{aligned}
\epsilon_{i t}&\sim \mathcal{N}_2\big([2; -2], [1 \  -0.7; -0.7 \  \ 1] \big), \\
\alpha_{ij}&=\bar{X}_{ij}'\beta_0/2+v_{ij}, \quad  \text{where} \ v_{ij}\sim \mathcal{N}(0, 1).
\end{aligned}
\end{equation*}

\item Design 3:  misspecified fixed effects
\begin{equation*}
\begin{aligned}
\epsilon_{ij t}&\sim \text{Gumbel}(0, 1),  \\
\alpha_{ij}&=(\bar{X}_{ij}/2-\bar{X}_{ik})'\beta_0*(1+v_{ij}), \quad  \text{where} \ v_{ij}\sim \mathcal{N}(0, 1).
\end{aligned}
\end{equation*}

\item Design 4:  misspecified distribution and misspecified fixed effects
\begin{equation*}
\begin{aligned}
\epsilon_{it}&\sim  \mathcal{N}_2\big([2; -2], [1 \  -0.7; -0.7 \  \ 1] \big), \\
\alpha_{ij}&=(\bar{X}_{ij}/2-\bar{X}_{ik})'\beta_0*(1+v_{ij}), \quad  \text{where} \ v_{ij}\sim \mathcal{N}(0, 1).
\end{aligned}
\end{equation*}

\end{itemize}

For the above four designs, I compare different approaches by reporting their root mean-squared error (rMSE) and mean of absolute deviation (MAD). For the complementarity parameter $\gamma_0$, I also report the probability of estimating the sign of substitution patterns incorrectly (Err) of the two estimators, defined as 
\[\text{Err}=E | \sign (Z_{i} \gamma_0)-\sign(Z_{i} \hat{\gamma} ) |.\]

 I study three different  sample sizes $N=\{1000, 2000, 4000\}$ and set the simulation repetitions to $B=500$. 
The performance of the estimator for $\gamma_0$ and $\beta_0$ is displayed in Table \ref{table:gamma} and in Table  \ref{table:beta}, respectively.

Table \ref{table:gamma} compares the performance of the two-step estimator with the parametric estimator for the complementarity parameter $\gamma_0$. The parametric estimator performs better only when its assumptions are correctly specified (design 1), but has a worse performance under misspecifications (designs 2-4) especially when both the parametric distribution and the model of fixed effects are misspecified.  The two-step estimator has uniform performance in all four designs, showing its advantage of performing robustly under different designs of parametric distributions and models of fixed effects. Moreover, as the sample size increases, the deviation and bias of the two-step estimator both shrink significantly. However, the bias of the parametric estimator does not decrease as the sample increases in designs 2-4, which shows the inconsistency of this estimator under misspecifications. 

Table \ref{table:beta} compares the performance of the two-step estimator with three other estimators described before for the coefficient $\beta_0$ under the four designs. Similarly, the two-step estimator performs uniformly better than the other three estimators in designs 2-4, and the difference becomes more significant as the sample size increases. To summarize, the results in Table \ref{table:gamma} and Table \ref{table:beta} demonstrate the advantage of the two-step estimator in performing robustly with respect to different parametric distributions or specifications of dependence structures between covariates and fixed effects.

% I compare the two-step estimator with the parametric estimator by reporting their standard deviation (SD), root mean-squared error (rMSE), and mean of absolute deviation (MAD). 
%\begin{equation*}
%\text{Err}=E | \sign (Z_{i} \gamma_0)-\sign(Z_{i} \hat{\gamma} ) |.
%\end{equation*}

% Let $\theta^k$ denote the $k$th element of the parameter $\theta$. The parameter $\theta_0$ can be only identified up to a constant, so the first element of is normalized to one: $\beta^1_0=1$ and the performance of the estimator of $\beta^2_0$ is displayed in Table \ref{table:beta}. 
%Since only the ratio $\tilde{\gamma}_0=\gamma_0^2/ \gamma_0^1$ matters for the substitution patterns, I focus on the results for the estimator of $\tilde{\gamma}_0$ in Table \ref{table:gamma}. 

\begin{table}[!htbp]
\centering
\caption{Performance Comparisons for $\hat{\gamma}$}
\label{table:gamma}
\begin{tabular}{cc |cccc|cccc}
\hline
\hline
 \multirow{2}{*}{$N$}&\multirow{2}{*}{Design}&\multicolumn{4}{c|}{Two-Step Est.}  &  \multicolumn{4}{c}{Parametric Est.}  \\
\cline{3-10} 
&  & Err & SD & rMSE & MAD  & Err & SD  & rMSE & MAD \\
\hline
\multirow{5}{*}{1000} & design 1  &0.025 & 0.263  &  0.315 & 0.267  & 0.007&  0.101 &0.101  &0.079  \\ [1ex]       
& design 2  &0.026 & 0.309 &  0.341 & 0.278  & 0.058 &  0.344 &0.741 &0.663  \\ [1ex]                                                                                                                                                                                                                                                                                                                                                                                                                                                                                                                                                                                                                                                                   
&design 3  & 0.023 & 0.287 &  0.306 &  0.251 &0.083 & 0.288 & 1.019 &0.977  \\[1ex]   
&design 4  & 0.025 &0.329 & 0.332 & 0.264 & 0.126 & 0.681 & 1.764 &1.628 \\[1ex]  
\hline
\multirow{5}{*}{2000} & design 1  &  0.022 & 0.258 & 0.281 & 0.229 &0.005 & 0.067 & 0.067 &  0.054  \\ [1ex]  
& design 2  & 0.022 &0.286 &0.294 &0.235 &  0.060&0.266 &0.719 &0.670  \\[1ex]   
& design 3  & 0.022 &0.295  &0.298 &0.234   &0.085 & 0.213 & 1.021  &0.998   \\[1ex]   
& design 4  &  0.021 &0.282 & 0.286 & 0.226 & 0.137 &0.580 &1.875 & 1.783    \\[1ex]  
\hline
\multirow{5}{*}{4000} & design 1  & 0.018 & 0.219  &  0.232 & 0.188  & 0.003 &  0.044 &0.044   &0.035  \\ [1ex]      
& design 2  & 0.021 & 0.270  &  0.270 & 0.223  & 0.060 &  0.195 &0.700  &0.672  \\ [1ex]                                                                                                                                                                                                                                                                                                                                                                                                                                                                                                                                                                                                                                                                    
&design 3  &  0.017 & 0.226 &  0.227 &  0.179 & 0.084 & 0.149 &0.983  & 0.972 \\[1ex]    
&design 4  & 0.020 & 0.250 & 0.271 & 0.218 & 0.141 & 0.437 & 1.873 & 1.822 \\[1ex]   
\hline
\end{tabular}
\end{table}

\begin{table}[!htbp]
\centering
\caption{Performance Comparisons for $\hat{\beta}$}
 \label{table:beta}
\centering
\begin{tabular}{cc |cc|cc|cc|cc}
\hline
\hline
& & \multicolumn{4}{c|}{Estimators with bundles} &  \multicolumn{4}{c}{Estimators assuming no bundles}\\  
\hline
 \multirow{2}{*}{$N$}&\multirow{2}{*}{Design}&\multicolumn{2}{c|}{Two-Step Est.} &  \multicolumn{2}{c|}{Parametric Est.}  & \multicolumn{2}{c|}{FE Logit Est.}  & \multicolumn{2}{c}{Semi. Est.} \\  
\cline{3-10}
 &  & rMSE & MAD & rMSE & MAD  & rMSE & MAD  & rMSE & MAD \\
 \hline
\multirow{5}{*}{1000} & design 1  &  0.126 & 0.101 & 0.072& 0.056  & 0.165 &  0.130 &0.155 & 0.131   \\[1ex]  
&design 2  &  0.125 & 0.101 &  0.432 & 0.389  &  0.243  & 0.187 & 0.161 & 0.134 \\[1ex]  
&design 3  &  0.128 & 0.100 &  0.304 & 0.279  &  0.168 & 0.131 &  0.149 &0.124 \\[1ex]  
&design 4 & 0.121 & 0.095 & 0.647 & 0.593 & 0.256 & 0.200 & 0.171 & 0.143 \\[1ex]  
\hline
\multirow{5}{*}{2000} & design 1  &  0.093 & 0.073 & 0.051 & 0.040  & 0.117 &  0.092 &0.114 & 0.092  \\ [1ex]  
&design 2  &  0.095 &0.065 & 0.114 & 0.095 &0.166 &  0.122 & 0.105 & 0.072\\ [1ex]  
&design 3  &0.094 &0.076 & 0.300 &0.286 &0.128 &0.101 & 0.115  &0.092 \\ [1ex]  
&design 4 & 0.099 &0.079 &0.612 &0.581  &0.205 &0.167 &0.132 & 0.109 \\ [1ex]  
\hline
\multirow{5}{*}{4000} & design 1  &  0.071 & 0.057 & 0.034 & 0.027 & 0.080 &  0.062 &0.079 & 0.061    \\ [1ex]  
&design 2  & 0.075 & 0.060&  0.385 & 0.368  &  0.157 & 0.135 & 0.091 & 0.072 \\[1ex]  
&design 3  &  0.070 & 0.048 & 0.203 & 0.198 & 0.076 & 0.050 & 0.075 & 0.045 \\[1ex]  
&design 4  & 0.077 & 0.063 & 0.606 & 0.588 & 0.168 & 0.144 & 0.104 &0.085\\[1ex]  \hline
\end{tabular}
\end{table}

\newpage

\section{Extension: Latent Complementarity} \label{sec:exte}

%The previous sections focused on the case where heterogeneity in the complementarity $\Gamma_{it}$ only comes from observed covariates: $\Gamma_{it}=\Gamma(Z_{i})$. This section accommodates unobserved heterogeneity in the complementarity across individuals. The latent complementarity term $\Gamma_{it}$ can be a random variable with an unknown distribution. 
%
%In this section, $\Gamma_{it}$ can be different for each individual regardless of their covariates, and it can be either positive or negative. The sign of $\Gamma_{it}$ captures the heterogeneous complementarity relationship among the two goods for each individual. Therefore, I focus on identifying the distribution of the sign of $\Gamma_{it}$ which represents the fraction of people for whom the two goods are complements or substitutes.  

The previous sections focused on the case where heterogeneity in the complementarity $\Gamma_{it}$ only comes from observed covariates: $\Gamma_{it}=\Gamma(Z_{i})$. This section accommodates unobserved heterogeneity in the complementarity across individuals by allowing $\Gamma_{it}$ to be different for each individual regardless of their covariates. The sign of $\Gamma_{it}$ captures the heterogeneous complementarity relationship among the two goods for each individual. Therefore, I focus on identifying the distribution of the sign of $\Gamma_{it}$ which represents the fraction of people for whom the two goods are complements or substitutes.

Next, I introduce some assumptions on the complementarity and error terms. 

\begin{ass}\label{ass:sta_gam}
The joint distribution of $(\epsilon_{it}, \Gamma_{it})$ conditional on $(\alpha_i,  X_{is}, X_{it})$ is stationary over time:
\begin{equation*}
(\epsilon_{is}, \Gamma_{is})\mid X_{is}, X_{it}, \alpha_i   \stackrel{d}{\sim}  (\epsilon_{it}, \Gamma_{it}) \mid X_{is}, X_{it}, \alpha_i \quad \text{for any} \ s, t\leq T.
\end{equation*}
\end{ass}

Assumption \ref{ass:sta_gam} is similar to Assumption \ref{ass:sta}, except it also assumes a stationarity condition for the complementarity $\Gamma_{it}$. Under the assumption that the complementarity $\Gamma_{it}$ only depends on observed covariates (Assumption \ref{ass:par}),  Assumption \ref{ass:sta_gam} degenerates to the stationarity condition in Assumption \ref{ass:sta} since $\Gamma_{it}$ is a constant conditional on the covariate. 
 
Assumption \ref{ass:sta_gam} only requires that the distribution of $\Gamma_{it}$ remains stationary over time, but it still allows the complementarity $\Gamma_{it}$ for each individual to vary over time. Moreover, this assumption does not restrict the dependence between the complementarity $\Gamma_{it}$ with the unobserved terms, including the fixed effects $\alpha_{i}$ and the error terms $\epsilon_{it}$.
 
 Let $X_i=(X_{it})_{t=1}^T$ collect the covariate of all time periods. 
\begin{ass}\label{ass:gam}
The complementarity $\Gamma_{it}$ is independent of the covariate $X_i$ conditional on the fixed effects $\alpha_i$: $\Gamma_{it}\independent X_i \mid \alpha_i$.
\end{ass}

Assumption \ref{ass:gam} assumes the independence between the complementarity $\Gamma_{it}$ and the vector of covariates for all periods. Under this condition, variation in all covariates can be used to identify the distribution of the complementarity $\Pr(\Gamma_{it}\geq 0)$. This assumption can be relaxed to accommodate the situation where there is a subset of covariates that are independent of $\Gamma_{it}$ while other covariates can be correlated with $\Gamma_{it}$. In such a scenario, the analysis can be conducted conditional on the covariates that are potentially correlated with the complementarity.  

Under the above assumptions, I establish the identification of the fraction of individuals for whom the two goods are complements, denoted as $\eta=\Pr(\Gamma_{it}\geq 0)$.  According to Assumption \ref{ass:sta_gam}, the distribution of $\Gamma_{it}$ is stationary over time so that $\eta$ does not depend on $t$. The identification result for $\Pr(\Gamma_{it}< 0)$ can be directly derived using the formula $\Pr(\Gamma_{it}<0)=1-\Pr(\Gamma_{it}\geq0)$ so it is omitted here.

The intuition of the identification strategy for $\eta$ is described as follows. The conditional demand for one good can be expressed as a mixture of two groups: people for whom the two goods are complements ($\Gamma_{it}\geq 0$) and people for whom the two goods are substitutes ($\Gamma_{it}<0$). When the covariate index of good $A$ increases, it will affect the demand for good $B$ for the two groups of people in different directions. This relationship can help identify the fraction of the two groups.

Similar to Section \ref{subsec:test}, the first step is to derive the sign of variation in covariate indices $(\Delta_{s, t} \delta_{A}, \Delta_{s, t} \delta_{B})$ from conditional choice probabilities. Let $\xi^{1}_{s,t} (x_s, x_t)$ and $\xi^{2}_{s,t} (x_s, x_t)$ be defined as
\begin{equation*}
\begin{aligned}
\xi_{s, t}^{1}(x_{s}, x_{t}) &= \mathbbm{1}\big\{P_s(\{j\} \mid x_{s}, x_{t})- P_t(\{j\} \mid x_{s}, x_{t})\geq 0, \  \forall j \in \{A, B, AB\}   \big\}, \\
\xi_{s, t}^{2}(x_{s}, x_{t})&=\mathbbm{1}\big\{P_s(\{j\} \mid x_{s}, x_{t})- P_t(\{j\} \mid x_{s}, x_{t})\geq 0,  \ \forall j \in \{A, AB, O\}  \big\}.
\end{aligned}
\end{equation*}

From the variation in observed choice probabilities, the sign of variation in covariate indices of two goods can be identified. When an increase in probabilities of all choices $\{A, B, AB\}$ is observed, it can be inferred that the covariate indices of both goods should increase. Similarly, an increase in probabilities of all choices except choice $B$ suggests that the covariate index of good $A$ increases and that of good $B$ decreases.

Let $\mathcal{X}^{1}_{s,t}=\{ (x_s, x_t)\mid \xi^{1}_{s,t} (x_s, x_t) =1\} $ and $\mathcal{X}^{2}_{s,t}=\{ (x_s, x_t)\mid \xi^{2}_{s,t} (x_s, x_t) =1\} $ denote the collection of covariates such that $\xi^{1}_{s,t} (x_s, x_t) =1$ and $\xi^{2}_{s,t} (x_s, x_t)=1$ respectively, which implies
\begin{equation*}
\begin{aligned}
 (x_s, x_t)\in \mathcal{X}^{1}_{s,t} \ \Longrightarrow  \ \Delta_{s,t} \delta_{A}\geq 0, \ \Delta_{s,t} \delta_{B}\geq 0, \\
(x_s, x_t)\in \mathcal{X}^{2}_{s,t}  \ \Longrightarrow \  \Delta_{s,t} \delta_{A}\geq 0, \ \Delta_{s,t} \delta_{B}\leq 0.
\end{aligned}
 \end{equation*}
 
Given the sign of covariate indices, using variation in the demand for two goods can identify the fraction of people for whom the two goods are complements $\eta=\Pr(\Gamma_{it}\geq 0)$. To convey the idea, I first consider that the covariate indices for goods $A$ and $B$ both increase: $(x_s, x_t)\in \mathcal{X}^{1}_{s,t}$. In this scenario, the demand for the two goods would increase for people for whom the two goods are complements, but may decline for people for whom the two goods are substitutes. Therefore, a decline in demand for either of the two goods in data can only come from people with $\Gamma_{it}<0$, which can help establish a lower bound for the fraction of people with $\Gamma_{it}<0$ and thus an upper bound for the fraction of people with $\Gamma_{it}\geq 0$. Similarly, a lower bound for $\eta=\Pr(\Gamma_{it}\geq 0)$ can be provided when covariates satisfy $(x_s, x_t)\in \mathcal{X}^{2}_{s,t}$.

The next proposition characterizes the partial identification results for $\eta=\Pr(\Gamma_{it}\geq 0)$.

\begin{prop}\label{hetero}
Under Assumptions \ref{ass:sta_gam}-\ref{ass:gam},  $\eta$ can be bounded as $\eta\in[L_{\eta}, U_{\eta}]$, where
\begin{equation*}
\begin{aligned}
L_{\eta}&=\sup_{(x_s, x_t) \in \mathcal{X}^{2}_{st}, \ell\in\{A,B\}, s,t\leq T } \Big\{(-1)^{\mathbbm{1}\{ \ell=A\} }[P_s(D_{\ell} \mid x_s, x_t)-P_t(D_{\ell} \mid x_s, x_t)] \Big\}, \\
U_{\eta}&=\inf_{(x_s, x_t)\in \mathcal{X}^{1}_{st}, \ell\in\{A,B\}, s,t\leq T } \Big\{P_s(D_{\ell} \mid x_s, x_t)-P_t(D_{\ell} \mid x_s, x_t) \Big\}+1. 
\end{aligned}
\end{equation*}

\end{prop}

\vspace{0.2cm}

Proposition \ref{hetero} establishes both lower and upper bounds for $\eta$ by exploiting variation in the demand for the two goods under different sets of covariate indices. According to the definition of the lower and upper bounds, it is always true that $L_{\eta}\geq 0, U_{\eta}\leq 1$ and the bounds use variation in all values of covariates over any two periods. The range of the bounds depends on variation in conditional demand of the two goods, and larger variation leads to tighter bounds.
 The result in Proposition \ref{hetero} also provides testable implications for Assumptions \ref{ass:sta_gam}-\ref{ass:gam} since it implies that the upper bound should be no smaller than the lower bound: $U_{\eta}\geq L_{\eta}$. The proof for Proposition \ref{hetero} is provided in Appendix \ref{proof:hetero}. 

The prior work by \cite{allen2022} also studies latent complementarity and provides bounds for the fraction of the population for whom the two goods are complements with cross-sectional data. They exploit an exclusion restriction and an independence assumption between the covariates and unobserved terms. My method complements their approach by considering panel data setting and using intertemporal variation over time. I allow covariates to be arbitrarily dependent with unobserved fixed effects. Moreover, I do not impose exclusion restrictions and can still partially identify $\eta$ when covariates of both goods change simultaneously. 

%\cite{allen2022} also study latent complementarity and provide bounds for the fraction of the population for whom the two goods are complements with cross-sectional data. Their paper exploits an exclusion restriction: there exists one covariate that only affects the utility of good $A$ but not good $B$. Also, their paper uses an independence assumption between the covariates and all unobserved terms.  As a complement to their paper, this paper considers panel data setting and mainly exploits intertemporal variation over time. My method allows covariates to be arbitrarily dependent with unobserved fixed effects. In addition, the analysis in this paper does not require an exclusion restriction and can still (partially) identify $\eta$ when covariates of both goods change simultaneously. 

\section{Conclusion}\label{sec:conc}

%This paper characterizes the sharp identification of a panel multinomial choice model allowing for bundles and provides testable implications for the substitution relationship between goods.
% The model in this paper allows for the possibility that two goods are either substitutes or complements and admits heterogeneous complementarity relationships through observed covariates. The identification analysis does not assume parametric distributions over idiosyncratic error terms and allows for endogeneity by admitting flexible dependence structures between observed covariates and unobserved fixed effects. 
 
This paper uses a panel multinomial choice model with bundles to study the substitution and complementarity relationship between goods.  The model imposes no parametric assumptions on the idiosyncratic error terms and allows for endogeneity by admitting flexible dependence structures between observed covariates and unobserved fixed effects. I provide testable implications for the substitution and complementarity relationship, and derive the sharp identification set for the model parameters.

The primary identification strategy is to derive identifying restrictions on unknown parameters through intertemporal variation in conditional choice probabilities that are identified from data. 
I construct conditional moment inequalities to characterize the identified set for estimation and inference. The method in the paper is shown via Monte Carlo simulations to perform more robustly than the parametric method concerning different specifications of fixed effects and distributions of error terms. In the extension, the paper allows for unobserved heterogeneity in the complementarity and establishes partial identification results. In the online Appendix, I also study the case of more than two goods, a class of nonseparable utility functions, as well as cross-sectional models.

This paper focuses on a static panel multinomial choice model where consumers' utility of goods only depends on characteristics in the same periods. It would be interesting to investigate how to identify the complementarity relationship in a dynamic model, where consumers' choices may also depend on past choices. The analysis would be more complicated as one has to disentangle the effect from previous choices and the current complementarity. In addition, it could also be worthwhile to explore how to identify the complementarity and utility coefficients with heterogeneous and unknown choice sets.

\bibliography{jmp}
\bibliographystyle{apalike}

\appendix

\section{Appendix} \label{sec:appe}

In the following proofs, I will suppress the covariate $Z_i$ and use $\Gamma_0$ to denote the incremental utility $\Gamma(Z_i)$.

\subsection{Proof of Lemma \ref{lem1}}
\begin{proof}
Lemma \ref{lem1} contains two results to be shown: $\Gamma_0\geq0 $ implies $s_{AB} \leq 0$, and $\Gamma_0 \leq0 $ implies $s_{AB}\geq 0$. I will show the proof for the first result, and the same idea can be applied to the second case. 

Suppose that the complementarity term is positive $\Gamma_0 \geq 0$, and I need to show $s_{AB}\leq 0$. From the definition of $s_{AB}$, proving $s_{AB}\leq 0$ is equivalent to showing that if $p_{Bs}>p_{Bt}$, then $\Pr(Y_{is}\in D_A\mid p_{Bs}, p_{Bt}, \tilde{x}) \leq \Pr(Y_{it}\in D_A\mid  p_{Bs}, p_{Bt}, \tilde{x}) $. The covariate $\tilde{x}$ is suppressed in this proof as it is fixed over time and it will not affect variation in conditional choice probabilities.

Let $\beta_{0,p}\leq 0$ denote the coefficient for  price $p_{\ell t}$. Let  $v_{i \ell  t}=\alpha_{i\ell }+\epsilon_{i\ell t}$ for $\ell\in \{A, B\}$.  The utility for good $B$ can be expressed as $u_{iB t}=p_{Bt}\beta_{0,p}+v_{iB t}$ and for good $A$ is  $u_{iAt}=v_{iAt}$ since all other covariates of good $A$ are suppressed. 

Let $ \mathcal{V}_{D_A}(p_{Bt})$ denote the collection of $v=(v_A, v_B)$ such that there exists one choice in $D_A=\{A, AB\}$ being chosen conditional on price $p_{Bt}$. The set $ \mathcal{V}_{D_A}(p_{Bt})$ includes two cases:  either choice $A$ or choice $AB$ has higher utility than all other options not in $D_A$. The set $\mathcal{V}_{D_A}(p_{Bt})$ can be expressed as follows:
\begin{equation*}
\begin{aligned}
 \mathcal{V}_{D_A}(p_{Bt})
 &=\big\{v  \mid  v_A \geq p_{Bt}\beta_{0,p}+v_{B}, \ v_{A}\geq 0\big \}\equiv \mathcal{V}_{1}(p_{Bt})  \\ 
 &\cup\big\{v \mid  v_{A}+\Gamma_0 \geq 0, \ v_{A}+ p_{Bt}\beta_{0,p}+v_{B}+\Gamma_0 \geq 0 \big \} \equiv  \mathcal{V}_{2}(p_{Bt}). \\
 \end{aligned}
 \end{equation*}
 
The demand for good $A$ given fixed effects and prices can be expressed as
\begin{equation*}
\Pr(Y_{it}\in D_A\mid  \alpha_i, p_{Bs}, p_{Bt})=\Pr( v_{it}\in \mathcal{V}_{D_A}(p_{Bt}) \mid \alpha_i, p_{Bs}, p_{Bt}).
\end{equation*}

Under Assumption \ref{ass:sta} (stationarity), the conditional distribution of $v_{it}$ is stationarity over time since the conditional distribution of $\epsilon_{it}$ is the same over time and the fixed effects $\alpha_i$ are constant.  Therefore, a larger set would imply a higher conditional probability as follows:
 \begin{equation*}
\mathcal{V}_{D_A}(p_{Bs})  \subseteq \mathcal{V}_{D_A}(p_{Bt}) \Longrightarrow \Pr(Y_{is}\in  D_A\mid \alpha_i, p_{Bs}, p_{Bt}) \leq \Pr(Y_{it}\in D_A\mid \alpha_i, p_{Bs}, p_{Bt}).
 \end{equation*}
 
By taking expectations with respect to the fixed effect $\alpha_i$ conditional on covariates, the above condition leads to
\begin{equation*}
\mathcal{V}_{D_A}(p_{Bs})  \subseteq \mathcal{V}_{D_A}(p_{Bt}) \Longrightarrow \Pr(Y_{is}\in D_A\mid p_{Bs}, p_{Bt}) \leq \Pr(Y_{it}\in D_A\mid p_{Bs}, p_{Bt}).
\end{equation*}
 
Proving Lemma \ref{lem1} is equivalent to showing that a higher price $p_{Bs}>p_{Bt}$ implies $\mathcal{V}_{D_A}(p_{Bs})  \subseteq \mathcal{V}_{D_A}(p_{Bt})$.  I will prove $\mathcal{V}_{D_A}(p_{Bs})  \subseteq \mathcal{V}_{D_A}(p_{Bt})$ by showing that for any element $v\in \mathcal{V}_{D_A}(p_{Bs}) $, it satisfies $v \in \mathcal{V}_{D_A}(p_{Bt})$. 
The proof will proceed by discussing two cases: $v\in \mathcal{V}_{1}(p_{Bs})$ and $v \in \mathcal{V}_{2}(p_{Bs}) $.

Case 1: $v \in \mathcal{V}_{1}(p_{Bs})$.  If $v$ satisfies $v_{A}\geq p_{Bt}\beta_{0, p}+v_{B}$ then $v\in \mathcal{V}_{1}(p_{Bt})$. Otherwise $v$ should satisfy
\begin{equation*}
 v_{A}< p_{Bt}\beta_{0, p}+v_{B}, \ v_{A}\geq 0.
\end{equation*}

Since $\Gamma_0 \geq 0$, it has the following implication:
\begin{equation*}
 v_{A}+\Gamma_0 \geq 0, \  (v_{A}+\Gamma_0)+ p_{Bt}\beta_{0, p}+v_{B} > (v_{A}+\Gamma_0)+v_A \geq 0.
\end{equation*}

Therefore we know that $v\in \mathcal{V}_{2}(p_{Bt}) \subseteq \mathcal{V}_{D_A}(p_{Bt})$.

Case 2:  $v\in \mathcal{V}_{2}(p_{Bs})$. According to the definition of the set $\mathcal{V}_{2}(p_{Bs})$, it  decreases with $p_{Bs}$ given  $\beta_{0,p}\leq 0$. Since $p_{Bs}>p_{Bt}$, it implies $v\in \mathcal{V}_{2}(p_{Bs}) \subseteq \mathcal{V}_{2}(p_{Bt})$. 

I have shown that for any element $v\in \mathcal{V}_{D_A}(p_{Bs}) $, it satisfies $v \in \mathcal{V}_{D_A}(p_{Bt})$ when $p_{Bs}>p_{Bt}$. Therefore, we can conclude that $\Gamma_0 \geq 0$ implies $s_{AB}\leq 0$.

\end{proof}

\subsection{Proof of Proposition \ref{prop1}}\label{proof:prop1}
\begin{proof}

%, $v_{iABt}=v_{iAt}+v_{iBt}+\Gamma_0$, and $v_{iOt}=0$. 
 Let  $v_{i\ell t}=\alpha_{i\ell}+\epsilon_{i\ell t}$ for $\ell \in \{A, B\}$ denote the sum of fixed effects and error term. For any set $K\subset \mathcal{C}$, let $\mathcal{V}_{K}(x_{t})$ denote the collection of $v=(v_A, v_B)$ such that there exists one choice in $K\subset \mathcal{C}$ being chosen given $X_{it}=x_t$. Let $v_{AB}=v_A+v_B+\Gamma_0$ and $v_O=0$ denote the error term for bundle $AB$ and the outside option respectively.  The set $\mathcal{V}_{K}(x_t)$ can be expressed as
 \begin{equation*}
 \mathcal{V}_{K}(x_{t})=\big\{v \mid \exists j\in K \ \text{s.t.} \  \delta_{jt}+v_{j}\geq \delta_{kt}+v_{k} \ \text{for all} \ k \in K^c    \big \},
 \end{equation*}
  where $\delta_{\ell t}=x_{\ell t}'\beta_0$ for $\ell\in \{A, B\}$, $\delta_{ABt}=\delta_{At}+\delta_{Bt}$, and $\delta_{Ot}=0$.
  
The conditional probability that there exists one choice in set $K$ being chosen can be expressed as follows:
 \begin{equation*}
\Pr(Y_{it}\in K\mid \alpha_i, x_s, x_t)=\Pr\big(v_{it} \in \mathcal{V}_{K}(x_t) \mid \alpha_i, x_s, x_t \big).
 \end{equation*}

 Under Assumption \ref{ass:sta} (stationarity), the conditional distribution of $v_{it}$ is stationarity over time for any $s\neq t$ since fixed effects $\alpha_i$ are constant over time.  Therefore, a larger set implies a larger conditional choice probability over time: 
\begin{equation}\label{eq:set}
\mathcal{V}_{K}(x_s) \subseteq \mathcal{V}_{K}(x_t) \ \Longrightarrow \  \Pr(Y_{is}\in K\mid \alpha_i, x_s, x_t) \leq \Pr(Y_{it}\in K\mid \alpha_i, x_s, x_t).  
\end{equation}

Next we will establish sufficient conditions on the parameter $\theta_0$ for the set relationship $\mathcal{V}_{K}(x_s) \subseteq \mathcal{V}_{K}(x_t)$, which would imply a decline in the conditional probability of the set $K$ over time. Then by contraposition, identifying restrictions for $\theta_0$ can be derived when there is an increase in choice probabilities over time.  Proposition \ref{prop1} comprises three parts of identifying restrictions, and the proof for each part is presented one by one.

\textbf{Part 1}: comparisons of the conditional probability of choice $j\in \mathcal{C}$ over time. According to the definition of the set $\mathcal{V}_{j}(x_{t})$, it  increases with respect to $\delta_{jt}-\delta_{kt}$ for $k\neq j$.  So when the covariate index of choice $j$ compared to all other choices decreases over time,  it implies the following set relationship:
\begin{equation*}
\delta_{js}-\delta_{ks} \leq \delta_{jt}-\delta_{kt} \ \forall  k\neq j \  \Longrightarrow  \   \mathcal{V}_{j}(x_s) \subseteq \mathcal{V}_{j}(x_t).
\end{equation*}

Note that the above relationship also holds for choice $AB$ with $\delta_{ABt}=\delta_{At}+\delta_{Bt}$. Plugging into the notation $\Delta_{s,t} \delta_{j}=\delta_{js}-\delta_{jt}$ and condition \eqref{eq:set}, it has the following implication:
\begin{equation*}
\Delta_{s,t} \delta_{j} - \Delta_{s,t} \delta_{k}\leq 0 \  \forall  k\neq j  \ \Longrightarrow  \   \Pr(Y_{is}= \{j\} \mid \alpha_i, x_s, x_t) \leq \Pr(Y_{it}=\{j\} \mid \alpha_i, x_s, x_t).
\end{equation*}

By contraposition and taking expectation over $\alpha_i$ conditional on  $(x_s, x_t)$, it yields the first identifying restriction in Proposition \ref{prop1}:
\begin{equation*}
 P_s(\{j\} \mid x_s, x_t) > P_t(\{j\} \mid x_s, x_t)
 \Longrightarrow \  \exists \ k\neq j  \ \text{s.t.}  \  \Delta_{s, t} \delta_{j}- \Delta_{s, t} \delta_{k}> 0.
 \end{equation*}

\textbf{Part 2}:  comparisons of the demand for good $\ell \in \{A, B\}$ over time.  I take good $A$ as an example to show the proof. The set $\mathcal{V}_{D_A}(x_t)$ can be expressed as the union of  two sets as follows: the set of choice $A$ and the set of choice $AB$ generating higher utility than other choices not in $D_A$, 
\begin{equation*}
\begin{aligned}
 \mathcal{V}_{D_A}(x_t)&=\big\{v  \mid \delta_{At}+v_A \geq  \delta_{Bt}+v_{B}, \  \delta_{At}+v_{A}\geq 0\big \}\equiv \mathcal{V}_{1}(x_t)  \\ 
 &\cup\big\{v \mid \delta_{At}+v_A+\Gamma_0  \geq 0, \delta_{At}+\delta_{Bt}+v_{AB}  \geq 0 \big \} \equiv \mathcal{V}_{2}(x_t).\\
 \end{aligned}
 \end{equation*}

To prove condition \eqref{id2} in Proposition \ref{prop1}, I look at the contrapositive statement of \eqref{id2} given as follows:
\begin{equation}\label{de}
\begin{aligned}
  \Big\{\Delta_{s,t} \delta_{A} \leq 0,  \ & \Delta_{s,t} (\delta_{A}+\sign(\Gamma_0) \delta_{B} ) \leq  0 \Big\} \lor  \big\{ |\Gamma_0|  \leq -\Delta_{s,t}\delta_{A} \}  \Longrightarrow   \mathcal{V}_{D_A}(x_{s})\subseteq  \mathcal{V}_{D_A}(x_{t}). %\Longrightarrow \Pr \big(Y_{is}\in  \mathcal{V}_{D_A}(x_{s})\mid \alpha_i, x_s, x_t \big)\leq \Pr\big(Y_{it}\in  \mathcal{V}_{D_A}(x_t)\mid \alpha_i, x_s, x_t \big)
\end{aligned}
\end{equation}

If the above condition is shown, then similarly condition \eqref{id2} is proved by contraposition and taking conditional expectation over  $\alpha_i$. The conditions on the parameter $\theta_0$  in \eqref{de} also depends on the sign of the complementarity $\Gamma_0$. 
I focus on the case $\Gamma_0> 0$ and the idea applies to the case $\Gamma_0 \leq 0$. 

When $\Gamma_0>0$,  the restriction on the parameter $\theta_0$ in \eqref{de} includes two parts: $C_1=\big\{\Delta_{s,t} \delta_{A} \leq 0 ,  \Delta_{s,t} (\delta_{A}+\delta_{B}) \leq  0 \big\} $ and $C_2=\{\Gamma_0\leq -\Delta_{s,t}\delta_{A} \}$.
Now I need to show that one of the two conditions $C_1 \lor C_2$ implies $ \mathcal{V}_{D_A}(x_s)\subseteq  \mathcal{V}_{D_A}(x_t)$. It can be proved by showing that any element $v$ belonging to $ \mathcal{V}_{D_A}(x_s)$ also belongs to $ \mathcal{V}_{D_A}(x_t)$ under either condition $C_1$ or $C_2$.  For any element $v\in \mathcal{V}_{D_A}(x_s)$,  I discuss two cases: $v \in \mathcal{V}_{1}(x_s) $ and $v \in \mathcal{V}_{2}(x_s)$.

\textit{Case 1}: $v \in \mathcal{V}_{1}(x_s) $.  If $v$ satisfies  $ \delta_{At}+v_{A}\geq \delta_{Bt}+v_{B}$, then $v\in \mathcal{V}_{1}(x_t)$ since either condition $C_1$ or $C_2$ both implies $\delta_{As} \leq \delta_{At}$. Otherwise $v$ should satisfy the following inequality:
\begin{equation*}
\delta_{Bt}+v_{B}> \delta_{At}+v_{A} \geq   \delta_{As}+v_{A}  \geq 0.
\end{equation*}

Since the complementarity is nonnegative $\Gamma_0 \geq 0$, the following conditions hold:
\begin{equation*}
\delta_{At}+v_{A}+\Gamma_0 \geq 0, \  (\delta_{At}+v_{A})+(\delta_{Bt}+v_{B})+\Gamma_0  \geq 0.
\end{equation*}

Therefore,  $v \in \mathcal{V}_{2}(x_t) \subseteq \mathcal{V}_{D_A}(x_t) $. 

\textit{Case 2}:  $v \in \mathcal{V}_{2}(x_s)$. I first consider that condition $C_1$ holds.
According to the definition of the set $\mathcal{V}_{2}(x_s)$, the set increases when the indices $\delta_{As}$ and $\delta_{As}+\delta_{Bs}$ both increase. Condition $C_1$ implies an increase in the covariate indices $\delta_{At}\geq \delta_{As}$ and $\delta_{At}+\delta_{Bt}\geq \delta_{As}+\delta_{Bs}$, so $v\in \mathcal{V}_{2}(x_t)$. 

Now consider that condition $C_2$ holds.  For any element $v \in \mathcal{V}_{2}(x_s)$, condition $C_2$ implies the following condition: 
\begin{equation*}
\delta_{At}+v_A\geq \delta_{As}+\Gamma_0+v_A \geq 0.
\end{equation*}

If  $v$ also satisfies the second condition in $\mathcal{V}_{2}(x_t)$ which is  $\delta_{At}+\delta_{Bt}+v_{A}+v_B+\Gamma_0 \geq 0$, then $v$ belongs to the set $\mathcal{V}_{2}(x_t)$: $v\in \mathcal{V}_{2}(x_t) $. Otherwise $v$ should satisfy
\begin{equation*}
\delta_{Bt}+v_B<-(\delta_{At}+v_{A}+ \Gamma_0)\leq \delta_{At}+v_A.
\end{equation*}

It implies that $v \in \mathcal{V}_{1}(x_t)$. I have shown whenever  $v \in \mathcal{V}_{D_A}(x_s)$, it satisfies $v \in \mathcal{V}_{D_A}(x_t)$ under either condition $C_1$ or $C_2$.

\textbf{Part 3}: comparisons of the sum of conditional probabilities of two choices over time. Condition \eqref{id3} includes two parts of identifying restrictions: one is the sum of the conditional probabilities of buying a single good and the other is the conditional probabilities of buying the bundle and the outside option. I focus on the condition using the sum of the conditional probabilities of buying a single good. 

 Similarly, I look at the contrapositive statement of condition \eqref{id3} in Proposition \ref{prop1}. Let $C_3=\{\min\big\{\Delta_{s,t} \delta_{A}, -\Delta_{s,t} \delta_{B}\big\} \leq  \Gamma_0\} $ and $C_4=\{ \Delta_{s, t}( \delta_{A}- \delta_{B}) \leq  0\}$, the contraposition of condition \eqref{id3} is given as
 \begin{equation*}
 C_3  \lor C_4  \  
 \Longrightarrow  \mathcal{V}_{A}(x_s) \subseteq \mathcal{V}_{\{A, AB, O\}}(x_t),
 \end{equation*}
where the set $\mathcal{V}_{A}(x_t)$ and $\mathcal{V}_{A, AB, O}(x_t)$ is given as
\begin{equation*}
\begin{aligned}
\mathcal{V}_{A}(x_t)&= \{v \mid \delta_{At}+v_A \geq 0, \  \delta_{At}+v_A \geq \delta_{Bt}+v_B, \  \delta_{Bt}+v_B+\Gamma_0 \leq 0 \}, \\
\mathcal{V}_{A, AB, O}(x_t)&=\{v\mid  \delta_{At}+v_A \geq \delta_{Bt}+v_B \ \text{or} \  \delta_{At}+v_A+\Gamma_0 \geq 0 \ \text{or} \ 0\geq \delta_{Bt}+v_B\}.
\end{aligned}
 \end{equation*}
 
First consider that condition $C_3$ holds, which has the following implications:
\begin{equation*}
\delta_{As} \leq \delta_{At}+\Gamma_0 \quad  \text{or} \quad  \delta_{Bs}+\Gamma_0  \geq \delta_{Bt}.
\end{equation*}

For any element $v\in \mathcal{V}_{A}(x_s)$, condition $C_3$ implies
\begin{equation*}
  \delta_{At}+v_A+\Gamma_0 \geq 0 \quad  \text{or} \quad  0\geq  \delta_{Bt}+v_B.
\end{equation*}

Therefore it is concluded that $v\in \mathcal{V}_{A, AB, O}(x_t)$.

When condition $C_4$ holds, it implies that $\delta_{As}-\delta_{Bs} \leq \delta_{At}-\delta_{Bt}$. Then the element $v$ satisfying $v\in \mathcal{V}_{A}(x_s)$ also satisfies $ \delta_{At}+v_A \geq \delta_{Bt}+v_B$,  we can conclude that $v\in \mathcal{V}_{A, AB, O}(x_t)$. 
The analysis for the sum of the conditional probabilities of purchasing the bundle and the outside option is similar, so I omit the analysis here.

\end{proof}

\subsection{Proof of Theorem~\ref{thm1}} \label{proof:thm1}
\begin{proof}
To prove sharpness, I need to show that for any parameter $\theta$ in the identified set $\Theta_I$, I can construct a data generating process such that it matches observed choice probabilities and satisfies assumptions.

Let $X_{i}=(X_{it})_{t=1}^{T}$ and $Y_i=(Y_{it})_{t=1}^T$ collect covariates and choice variables at all periods.  Let $F_{Y\mid X}(j_1, j_2,..., j_T\mid x)$ denote joint choice probabilities of choosing $j_t\in \mathcal{C}$ at all periods $t\leq T$ given $X_i=x$, which are identified from data. I set  fixed effects to be zero $\alpha_i=0$ and focus on constructing the conditional distribution of the error term $\epsilon_i\mid x$.

The first requirement of sharpness is that the constructed distribution of error terms can match the observed choice probabilities $F_{Y\mid X}(j_1, j_2,..., j_T\mid x)$ according to the utility function of all choices:
\begin{equation}\label{req1}
F_{Y\mid X }(j_1, j_2,..., j_T\mid x)=\Pr(u_{i j_t t} \geq u_{i k_t t} \quad \forall k_t\neq j_t, \forall t\leq T\mid  x).
\end{equation}
The left hand term represents observed choice probabilities in data, and the right hand term represents choice probabilities generated from the model which depend on the constructed distribution of the error term.

The second requirement is that the constructed distribution of the error term satisfies Assumption \ref{ass:sta} (stationarity), which is equivalent to the following condition given $(X_{is}, X_{it})=(x_s, x_t)$:
\begin{equation}\label{req2}
\Pr(\epsilon_{i s} \in K \mid x_{s}, x_{t})=\Pr(\epsilon_{it} \in K \mid x_{s}, x_{t}) \quad \text{for any set} \ K.
\end{equation}

To construct a conditional distribution of the error term to satisfy the above two requirements, the first step is to construct choice sets. Since only consumers' choices are observed in data, I will define choice sets that determine observed choices and construct the conditional distribution of $\epsilon_i \mid x$ over those sets. The distribution over other sets does not matter for observed variables and can be constructed arbitrarily as long as it satisfies the assumptions. 
 Let $\mathcal{E}_{K}(x_{t})$ denote the collection of $\epsilon=(\epsilon_A, \epsilon_B)$ such that there exists one choice in the set $K$ being selected given $X_{it}=x_t$, defined as
\begin{equation*}
\mathcal{E}_K(x_t)=\{\epsilon \mid \exists j\in K \  \text{s.t.} \ \delta_{jt}+\epsilon_{j} \geq \delta_{kt}+\epsilon_k \quad \forall k\in  K^c \mid x_t \},
\end{equation*}
where $\epsilon_{AB}=\epsilon_A+\epsilon_B+\Gamma_0$ and $\epsilon_O=0$. When $K=\{j\}$ is a singleton, $\mathcal{E}_j(x_t)$ is the set of error terms such that choice $j$ is selected given $x_t$.

The four choice sets $\{ \mathcal{E}_{j}(x_t)\}_{j\in \mathcal{C}}$ form partitions of the space of $\epsilon_{it}$ conditional on $x_t$.   
The conditional probability of selecting choice $j$ can be represented as follows:
\begin{equation*}
\Pr(Y_{it}=j\mid x_t)=\Pr(\epsilon_{it} \in \mathcal{E}_j(x_t)\mid x_t).
\end{equation*}

For any $j_t\in \mathcal{C}$, the first requirement (condition \eqref{req1}) is satisfied when we construct the conditional distribution of $\epsilon_i\mid x$ on the set $\mathcal{E}_{j}(x_t)$ as follows:
\begin{equation}\label{req3}
F_{Y \mid X}(j_1, j_2,..., j_T\mid x)=\Pr(\epsilon_{i1}\in \mathcal{E}_{j_1}(x_1),...,\epsilon_{iT}\in \mathcal{E}_{j_T}(x_T)\mid x).
\end{equation}

The joint distribution of $\epsilon_i \mid x$ over choice sets $ \mathcal{E}_{j}(x_t)$ is pinned down to match observed choice probabilities. Now I only need to verify that the stationarity assumption in condition \eqref{req2} can be satisfied. To show it, I will construct a marginal distribution of $\epsilon_{it}\mid (x_{s}, x_{t})$ over smaller sets such that it is stationary over any two periods $s\neq t$ and it is consistent with the distribution over choice sets derived from equation \eqref{req3}. 

Equation \eqref{req3} restricts the marginal distribution of $\epsilon_{it}\mid (x_{s}, x_{t})$ on the choice set $\mathcal{E}_{j}(x_t)$ at each $t\leq T$. The choice set $\mathcal{E}_{j}(x_t)$ depends on $x_t$ so it changes over time when the covariate $x_t$ changes. It is difficult to compare the two distributions defined over different sets at different periods and verify the stationarity assumption. To tackle this difficulty, I construct the marginal distribution of $\epsilon_{is}\mid (x_{s}, x_{t})$ and $\epsilon_{it}\mid (x_{s}, x_{t})$ on the intersection of the two choice sets $\mathcal{E}_{j}(x_s)$ and $\mathcal{E}_{j}(x_t)$, then their distributions are defined over the same set. Let $J_{j,k}(x_s, x_t)$ denote the intersection of the two sets $\mathcal{E}_{j}(x_s)$ and $\mathcal{E}_{j}(x_t)$, defined as follows:
\begin{equation*}
J_{j, k}(x_s, x_t)=\mathcal{E}_j(x_s) \cap \mathcal{E}_k(x_t).
\end{equation*}

Let $P_t(j\mid x_s, x_t)=\Pr(Y_{it}=j\mid x_s, x_t)$ denote the marginal probability of choosing $j$ at time $t$ which is identified from data. The requirements for the conditional distribution of $\epsilon_{it}\mid (x_{s}, x_{t})$ over the set $J_{j,k}(x_s, x_t)$ are equivalent to the following conditions: for any $j, k\in \mathcal{C}$ and any $s\neq t$,
\begin{equation} \label{req4}
\begin{aligned}
\Pr(\epsilon_{is}\in J_{j, k}(x_s, x_t) \mid x_s, x_t)&=\Pr(\epsilon_{it}\in J_{j, k}(x_s, x_t) \mid x_s, x_t), \\ 
\sum_k \Pr(\epsilon_{is}\in J_{j, k}(x_s, x_t) \mid x_s, x_t)&=P_s(j\mid x_s, x_t), \\
\sum_j \Pr(\epsilon_{it}\in J_{j, k}(x_s, x_t) \mid x_s, x_t)&=P_t(k\mid x_s, x_t).   \\
\end{aligned}
\end{equation}

The first equation guarantees the conditional stationarity assumption (Assumption \ref{ass:sta}), and the other two equations ensure that the constructed marginal distribution of $\epsilon_{it}\mid (x_s, x_t)$ is consistent with observed choice probabilities.  

Now I need to show that there exists nonnegative probabilities of $\epsilon_{it}\mid (x_{s}, x_{t})$ over the set $J_{j, k}(x_s, x_t)$ such that condition \eqref{req4} holds.
Let $r_{j,k}(x_s, x_t)\geq 0$ denote the conditional probability over the set $J_{j, k}$:
\begin{equation*}\
r_{j,k}(x_s, x_t)=\Pr(\epsilon_{is}\in J_{j, k}(x_s, x_t) \mid x_s, x_t)=\Pr(\epsilon_{it}\in J_{j, k}(x_s, x_t) \mid x_s, x_t).
\end{equation*}

The stationarity assumption is satisfied since the probability  $r_{j,k}(x_s, x_t)$ is time invariant. 
For the following analysis, I will suppress the covariate $(x_s, x_t)$ for the conditional probabilities $r_{j,k}(x_s, x_t)$ and $P_t(j \mid x_s, x_t)$ to simplify notation. I only need to construct nonnegative probabilities $r_{j,k}\geq 0$ such that the last two conditions in \eqref{req4} hold for all $j, k\in \mathcal{C}$:
\begin{equation}\label{req5}
\begin{aligned}
\sum_k r_{j, k}=P_s(j), \\
\sum_j r_{j, k}=P_t(k).
\end{aligned}
\end{equation}

This proof focuses on the case $\Gamma_0 \geq 0$ and the idea applies to the case $\Gamma_0<0$. The construction of $r_{j, k}$ depends on the relationship between  covariate indices and the complementarity term
 $\{ \Delta_{s,t}\delta_{A}, \Delta_{s,t} \delta_{B}, \Delta_{s, t}\delta_{AB}, \Gamma_0 \}$, since their relationship determines the relationship between choice sets. I discuss the following cases to show the construction of $r_{j, k}$.

\textbf{Case 1}: $\Delta_{s,t} \delta_{A}\geq \Delta_{s,t} \delta_{AB}\geq 0 \geq  \Delta_{s,t}\delta_{B}$, and $\Gamma_0\geq  \min \{ \Delta_{s,t}\delta_{A}, - \Delta_{s,t}\delta_{B} \} $. From the proof for Proposition~\ref{prop1} in \ref{proof:prop1}, the relationship between covariate indices and the complementarity term implies the following set inclusion relationship:
\begin{equation*}
\begin{aligned}
\mathcal{E}_J (x_t) &\subseteq  \mathcal{E}_J(x_s) \quad  \text{for any} \ J\in \{ \{A\}, \{A, AB\}, \{A, AB, O\}  \}, \\
\mathcal{E}_B(x_t) &\subseteq  \mathcal{E}_{\{B, AB, O\} }(x_s).
\end{aligned}
\end{equation*} 

According to the definition of $J_{j, k}$, the above set inclusion relationship implies that the following sets are empty:
\begin{equation*}
J_{k_1, A}=J_{k_2, AB}=J_{B, O}=J_{A, B}=\emptyset \quad \text{for} \ k_1\neq A, k_2=\{B, O\}.
\end{equation*}

Given the relationship between covariate indices and the complementarity term, the contraposition of conditions \eqref{id1}-\eqref{id3} in Proposition~\ref{prop1}  are equivalent to the following inequalities:
\begin{equation} \label{req6}
\begin{aligned} 
P_{t}(A) &\leq P_s(A), \\
P_t(B) &\geq P_s(B), \\
P_t(A)+P_t(AB) &\leq P_s(A)+P_s(AB), \\
P_{t}(B)+P_s(A) &\leq 1.
\end{aligned}
\end{equation}

Now I need to show that when the above restrictions \eqref{req6} hold, there exists nonnegative probabilities $r_{j, k}\geq0$ on nonempty sets $J_{j,k}$ such that \eqref{req5} holds. 
%
%The following displays all probabilities $r_{j, k}$ which are not determined:
%\begin{equation*}
%\begin{aligned}
%&r_{B,B}                   \\
%&r_{O,B}    &     &r_{O, O}      \\
%&r_{AB,B}  &    &r_{AB, O}  &   &r_{AB, AB}  \\
%&                &    &r_{A, O}    &   &r_{A, AB}  &   &r_{A, A} \\ 
%\end{aligned}
%\end{equation*}
%

%Condition \eqref{req5} requires that the sum of each row of $r_{j, k}$ equals to $P_s(j)$ and the sum of each column equals to $P_t(j)$. Then the following two probabilities can be determined:
%\begin{equation*}
%r_{B, B}=P_s(B),  \qquad  r_{A, A}=P_t(A).
%\end{equation*}
%
%I look at the sum of probabilities in the first column and second row:
%\begin{equation*}
%\begin{aligned}
%&r_{O, B}+r_{AB, B}=P_t(B)-P_s(B), \\
%&r_{O, B}+r_{O, O}=P_s(O).
%\end{aligned}
%\end{equation*}
%Based on the above conditions, I construct nonnegative probabilities as follows:
%\begin{equation*}
%\begin{aligned} 
%&r_{O, B}=\min\{P_t(B)-P_s(B),  P_s(O)  \}, \\
%&r_{AB, B}=P_t(B)-P_s(B)-r_{O,B},  \\
%&r_{O,O}=P_s(O)-P_{O, B}.
%\end{aligned}
%\end{equation*}
%
%Similarly, I look at the sum of probabilities in the last row and third column, and the corresponding probabilities can be constructed as
%\begin{equation*}
%\begin{aligned}
%&r_{A, AB}=\min\{P_s(A)-P_t(A),  P_t(AB) \}, \\
%&r_{A, O}=P_s(A)-P_t(A)-r_{AB, A}, \\
%&r_{AB, AB}=P_t(AB)-r_{AB, A}.
%\end{aligned}
%\end{equation*}

The probabilities on nonempty sets $J_{j,k}$ are constructed as follows:
\begin{equation*}
\begin{aligned} 
&r_{O, B}=\min\{P_t(B)-P_s(B),  P_s(O)  \}, \quad r_{AB, B}=P_t(B)-P_s(B)-r_{O,B}, \\
&r_{O,O}=P_s(O)-P_{O, B}, \quad r_{A, AB}=\min\{P_s(A)-P_t(A),  P_t(AB) \},\\
&r_{A, O}=P_s(A)-P_t(A)-r_{AB, A}, \quad r_{AB, AB}=P_t(AB)-r_{AB, A}.
\end{aligned}
\end{equation*}

The last probability $r_{AB, O}$ can be constructed as follows: 
\begin{equation*}
r_{AB, O}=
\left\{
\begin{aligned}
& 1-P_{t}(B)-P_s(A) \                  &\text{if} \ P_s(A) \geq P_t(\{A, AB\}),  P_t(B) \geq P_s(\{B, O\}),  \\
&P_s(AB) \                                 &\text{if} \ P_s(A) \geq P_t(\{A, AB\}),  P_t(B)\leq P_s(\{B, O\}),  \\
&P_t(O)  \                                    &\text{if} \ P_s(A)\leq P_t(\{A, AB\}),    P_t(B) \geq P_s(\{B, O\}),  \\
&P_s(\{A, AB\})-P_t(\{A, AB\})  \  &\text{if} \ P_s(A)\leq P_t(\{A, AB\}),    P_t(B)\leq P_s(\{B, O\}).
\end{aligned}
\right.
\end{equation*}

It can be verified that all constructed probabilities are nonnegative by their definition, and the probability $r_{AB, O}$ is nonnegative under condition \eqref{req6}.
The idea of constructing nonnegative probabilities $r_{j, k}$ for the following cases is similar.

\textbf{Case 2}:  $\Delta_{s,t} \delta_{A}\geq \Delta_{s,t} \delta_{AB}\geq  0\geq  \Delta_{s,t} \delta_{B}$ and $ \Gamma_0< \min\{\Delta_{s, t} \delta_A,  -\Delta_{s, t} \delta_B\}$. It implies following set inclusion relationship:
\begin{equation*}
J_{k_1, A}=J_{k_2, AB}=J_{k_3, O}=\emptyset \quad  \text{for} \ k_1\neq A, k_2=\{B, O\}, k_3=\{B, AB\}.
\end{equation*}

Given the relationship between covariate indices and the complementarity term, the contraposition of conditions in Proposition \ref{prop1} implies the following inequalities:
\begin{equation} \label{req7}
\begin{aligned} 
P_{t}(A) &\leq P_s(A), \\
P_t(B) &\geq P_s(B), \\
P_t(A)+P_t(AB) &\leq P_s(A)+P_s(AB), \\
P_{t}(B)+P_t(AB) &\geq P_s(B)+P_s(AB).
\end{aligned}
\end{equation}

The probability $r_{j, k}$ can be constructed as follows:
\begin{equation*}
\begin{aligned}
&r_{B, B}=P_s(B),             &                    &r_{A, A}=P_t(A),   \\
&r_{O, O}=\min\{P_t(O), P_s(O)\},     &     &r_{O, B}=P_s(O)-r_{O, O},    &      &r_{A, O}=P_t(O)-r_{O, O},\\
&r_{AB, AB}=\min\{P_t(AB), P_s(AB)\}, &  &r_{AB, B}=P_s(AB)-r_{AB, AB}, &  &r_{A, AB}=P_t(AB)-r_{AB, AB}. \\
\end{aligned}
\end{equation*}

By construction, the above probabilities are all nonnegative. The last probability is determined as $r_{A,B}=P_s(A)+P_t(B)-1+r_{AB, AB}+r_{O,O} $, and  it can be shown to be nonnegative $r_{A, B}\geq 0$ when condition \eqref{req7} holds.

The analysis for remaining cases are similar except exchanging the order of $A$ and $B$ or exchanging $AB$ and $O$.

%\textbf{Case 3}: $\Delta_{s,t} \delta_{A}\geq  0 \geq \Delta_{s,t} \delta_{AB}\geq  \Delta_{s,t} \delta_{B}$. In this case, I also need to discuss two cases $ \Gamma_0\geq  \min\{\Delta_{s, t} \delta_A,  -\Delta_{s, t} \delta_B\}$ and $ \Gamma_0< \min\{\Delta_{s, t} \delta_A,  -\Delta_{s, t} \delta_B\}$.  The two cases are similar to case 1 and case 2 except switching the place of bundle $AB$ and  outside option $O$. 
%
%\textbf{Case 4}: $\Delta_{s,t} \delta_{AB}\geq \Delta_{s,t} \delta_{A}\geq  \Delta_{s,t} \delta_{B}\geq 0$. The construction is the same with case 2, except changing the place of  $AB$ with  $A$ and changing  $B$ with  $O$.
%
%\textbf{Case 5}: $0\geq \Delta_{s,t} \delta_{AB}\geq \Delta_{s,t} \delta_{A}\geq \Delta_{s,t} \delta_{B}$. The construction is the same with case 3 by exchanging the order of the bundle $AB$ and the outside option $O$. 
%
%\textbf{Case 6}: all other cases are the same as above cases, except exchanging the place of choice $A$ and choice $B$.

\end{proof}

\subsection{Proof of Theorem \ref{thm2}}
\begin{proof}
I first show the proof for point identification of the coefficient $\beta_0$ and the idea is similar to the parameter $\gamma_0$. The first step is to show that for any candidate $b \neq k \beta_0$, there exists some value of the covariate such that the sign of the covariate index $\Delta X_{i\ell}'\beta$ for good $\ell\in\{A, B\}$ is different under the true parameter $\beta_0$ and the candidate $b$. I take good $A$ as an example to illustrate the idea.  
 
From Assumption \ref{ass:x}, the conditional density of $\Delta X_{iA}^{k_A}$ is positive everywhere. Let $\Delta \tilde{X}_{iA}=\Delta X_{iA}\setminus \Delta X_{iA}^{k_A}$ denote the remaining covariates in $\Delta X_{iA}$ and $\tilde{\beta}_0$ denote its coefficient.  Consider that the coefficient of $\Delta X_{iA}^{k_A}$ is positive $\beta_0^{k_A}>0$, and the analysis applies to the case $\beta_0^{k_A}<0$.  For any candidate $b$, I discuss three cases: $b^{k_A}<0$, $b^{k_A}=0$, and $b^{k_A}>0$.

Case 1: $b^{k_A}<0$. When the covariate $\Delta X_{iA}^{k_A}$ takes a large positive value $\Delta X_{iA}^{k_A}=\Delta x_{A}^{k_A}\rightarrow +\infty$ and all other covariates take a bounded number in their support, it implies that $\Delta x_{A}'\beta_0>0$ and $\Delta x_{A}'b<0$ since the true coefficient and the candidate coefficient have different signs $\beta_0^{k_A}>0$ and $b^{k_A}<0$.

Case 2: $b^{k_A}=0$. For any value $\Delta X_{iA}=\Delta x_{A}$, the value of $\Delta x_{A}'b$ is either positive or nonpositive. First consider that $\Delta x_{A}'b>0$ is positive. When $\Delta x_{A}^{k_A}$ takes a large negative number $\Delta x_{A}^{k_A}\rightarrow -\infty$ such that $\Delta x_{A}'\beta_0<0$, which has a different sign from $\Delta x_{A}'b$. Similarly, if $\Delta x_{A}'b\leq 0$, there exists $\Delta x_{A}^{k_A}\rightarrow +\infty$ such that $\Delta x_{A}'\beta_0>0$.
                                                                                                                                                                                                                                                                                                                                                                                                                                                                                                                                                                                
Case 3: $b^{k_A}>0$. Assumption \ref{ass:x} says that the support of the covariate $\Delta X_{iA}$ is not contained in any proper linear subspace, so there exists $\Delta \tilde{X}_{iA}=\Delta \tilde{x}_A$ such that $\Delta \tilde{x}_{A}' \tilde{\beta}_0 / \beta_0^{k_A} \neq \Delta \tilde{x}_{A}'  \tilde{b} /b^{k_A}$. Suppose that $ \Delta \tilde{x}_{A}'\tilde{\beta}_0 /\beta_0^{k_A}-\Delta \tilde{x}_{A}'  \tilde{b} /b^{k_A}=k>0 $, then when the covariate takes the value $\Delta X_{iA}^{k_A} =-\Delta \tilde{x}_{A}'  \tilde{b} /b^{k_A}-\epsilon$ with $0<\epsilon<k$. The sign of the covariate index satisfies: $\Delta x_{A}'\beta_0=\beta_0^{k_A}(k-\epsilon)>0$ and $\Delta x_{A}' b=-b^{k_A}\epsilon<0$. The construction is similar when $k<0$.

I have shown that there exists some value of the covariate such that the sign of the index $\Delta X_{i\ell}'\beta$ is different under the true parameter $\beta_0$ and the candidate parameter $b$. From Assumption \ref{ass:ep}, there exists at least one choice such that the conditional probability of this choice changes in strictly different directions under $\beta_0$ and $b$ so $\beta_0$ is identified. For example, if $\Delta x_{\ell}' \beta_0>0 $ and $\Delta x_{\ell}' b\leq 0$ for $\ell \in \{A, B\}$, then the conditional probability of choosing $AB$ strictly increases under $\beta_0$ and decreases under $b$. The analysis for other cases is similar.

Now we look at the proof for the parameter $\gamma_0$, which is similar to the above analysis. The sign of variation in covariate index $\Delta x_{\ell}'\beta_0$ is identified for $\ell \in \{A, B\}$ since $\beta_0$ is point identified. For any $\gamma\neq k \gamma_0$, we know that there exists some value of covariate $Z_i=z$ such that the sign of the complementarity $z'\gamma_0\neq 0$ is different from the sign of $z'\gamma$ under the large support assumption of $Z_i$ (Assumption \ref{ass:z}). Then, the conditional demand of good $A$ or good $B$ will change in strictly different directions under $\gamma_0$ and $\gamma$ given some value of covariates. For example, consider that the covariate of good $A$ is fixed over time and the covariate of good $B$ increases $\Delta x_B'\beta_0>0$. If $z'\gamma_0>0$ and $z'\gamma\leq 0$, then the demand for good $A$ will strictly increase under $\gamma_0$ but decrease under $\gamma$ by Condition \ref{id2} in Proposition \ref{prop1}. The opposite holds for the other case $z'\gamma_0<0$ and $z'\gamma \geq 0$. Therefore, the parameter $\gamma_0$ is point identified.

\end{proof}

\newpage

\section{Online Appendix}

\subsection{Extension: More Than Two Goods} \label{sec:more}

Although there are numerous applications where we are interested in the substitution relationship between two goods, it would be interesting to explore how the approach can be extended to accommodate more than two goods. In this section, I allow any finite number of goods $\mathcal{L}=\{A, B,..., J\}$. The letter $J$ is used to represent the last good, and any other letter can be used to incorporate additional goods. Let $\Gamma_{ibt}$ denote the incremental utility from choosing bundle $b$ compared to the sum of utilities of a single good $\ell \in \mathcal{L}$ in $b$:
\[\Gamma_{ibt}=u_{ibt}-\sum_{\ell \in b} u_{i\ell t}. \]

I consider the same assumption for $\Gamma_{ibt}$ as Assumption \ref{ass:par} where the incremental utility only depends on observed covariate $Z_i$: 
\[ \Gamma_{ibt}=\Gamma_b(Z_i). \]

Function $\Gamma_b$ is allowed to vary across bundles, which permits heterogeneous incremental utilities for different bundles. In this Appendix, the covariate $Z_i$ is suppressed so I use $\Gamma_b$ to represent the incremental utility for bundle $b$.

%I consider the incremental utility $\Gamma_{ib}=\Gamma_{b}$ of bundle $b$ to be constant across agents, and the analysis can be extended to the case where $\Gamma_{ib}$ depends on observed covariates as in Assumption \ref{ass:par}. 

The analysis becomes more complex when generalizing beyond two goods, posing several challenges. The feasible choices increase exponentially when allowing for all possible combinations of goods. The incremental utilities of all bundles compared to buying a single good could be heterogeneous across bundles, and the sign of them can be either positive or negative.
The demand for a good depends on covariate indices of all goods and the incremental utilities of all possible bundles. 
In this scenario, the relationship between the demand for a good and the incremental utilities of all possible bundles as well as all covariate indices becomes intractable.

Due to those challenges, I consider two assumptions to simplify the analysis with more than two goods. The first condition requires that consumers can purchase at most two goods in one period. By making this assumption, bundles of more than two goods are eliminated, which reduces the number of feasible choices available to consumers. 
This assumption may be justifiable as consumers could be subject to budget or inventory limitations that prevent them from purchasing more than two goods simultaneously.
 Under this condition, consumers' choice set is simplified as $\mathcal{C}=\{O, j, j_1 j_2 \ \forall  j\in \mathcal{L}, j_1\neq j_2 \in \mathcal{L}\}$. 
Second, I consider that we focus on the substitution relationship between goods $A$ and $B$, and allow its incremental utility $\Gamma_{AB}$ to be positive or negative. The incremental utility of all other bundles is assumed to be negative: $\Gamma_{j_1 j_2}\leq 0$ for $j_1\neq j_2$ and $j_1j_2\neq AB$. This condition still allows for the potential purchase of bundle $j_1 j_2$,  while restricting only the sign of its incremental utility. 

Under the above two assumptions, the following proposition characterizes the identifying restrictions for $(\beta_0, \Gamma_{AB})$.

\begin{prop} \label{prop:three}
Under Assumption \ref{ass:par}-\ref{ass:sta}, the following conditions hold: for any $s\neq t\leq T$, 

\begin{itemize}

\item comparisons of CCP of good $j\in \mathcal{L}$: 
\begin{equation*}
\begin{aligned}
P_s(\{j\}\mid x_s, x_t)>  P_t(\{j\} \mid x_s, x_t)  \Longrightarrow   \{ \Delta_{s,t} \delta_{j}>0 \} \lor \{ \exists k \neq j \in \mathcal{L}, \  \Delta_{s,t} \delta_{k}<0\};
\end{aligned}
\end{equation*}

\item comparisons of CCP of bundle $j_1j_2$ for $j_1\neq j_2\in \mathcal{L}$:
\begin{equation*}
\begin{aligned}
P_s( \{j_1j_2 \} \mid x_s, x_t) > & P_t(\{j_1j_2 \} \mid x_s, x_t)  \Longrightarrow   \\
&\big\{ \exists j\in \{j_1, j_2\}, \ \Delta_{s,t} \delta_{j}>0 \big\} \lor    \{ \exists  k\notin \{j_1, j_2\}, \ \Delta_{s,t} \delta_{k}<0\};
\end{aligned}
\end{equation*}

\item comparisons of  CCP of $D_\ell= \{\ell, AB\}$ for $\ell\in \{A, B\}$:
\begin{equation*}
\begin{aligned}
 P_s(D_{\ell} \mid x_s, & x_t) >  P_t(D_{\ell} \mid x_s, x_t) \Longrightarrow  \{ \Delta_{s,t}\delta_{\ell}> 0\} \lor \{ \Delta_{s,t} (\delta_{\ell}+\sign(\Gamma_{AB})\delta_{\ell_{-1}})> 0\} \\
&  \lor \{ \exists k\in \{C,..,J\}, \ \Delta_{s, t}(\delta_{\ell_{-1}}-\delta_{k})> 0 \} \lor \{ \exists k\in \{C,..,J\}, \ \Delta_{s,t}\delta_k < 0\},
\end{aligned}
\end{equation*}
where $\ell_{-1} \in \{A, B\} $ and $\ell_{-1}\neq \ell$.

\end{itemize}

\end{prop}

Proposition \ref{prop:three} extends the results for two goods to allow for any finite number of  goods. It still derives identifying restrictions on the utility coefficient $\beta_0$ and the incremental utility $\Gamma_{AB}$ from intertemporal variation in conditional choice probabilities. The identifying conditions are more complex due to the larger number of feasible choices to consumers and the need to consider variation in covariate indices of all goods.

The first part exploits variation in buying a single good to identify the utility coefficient $\beta_0$. When we observe an increase in the probability of buying a single good, e.g., good $A$, it can be inferred that either the covariate index of good $A$ increases or that of some other goods decreases. Similarly, the second part looks at variation in purchasing a bundle over time. If the probability of selecting a bundle (such as $AB$) increases, it indicates either an increase in the covariate index of a good (either $A$ or $B$) within that bundle or a decrease in a good outside of the bundle.

The last part in Proposition \ref{prop:three} uses variation in the conditional probability of choosing the set $D_A=\{A, AB\}$ (or $D_B$) to infer the sign of the incremental utility $\Gamma_{AB}$. To convey the idea, I first consider that the covariates of all other goods $\{C,...,J\}$ are fixed. If goods A and B both improve but the probability of purchasing $\{A, AB\}$ declines, it suggests that some individuals switch to buying only good $B$ instead of  bundle $AB$ and the incremental utility is negative $\Gamma_{AB}<0$.  When covariates of other goods $\{C,...,J\}$ can also change, then an increase in the probability of selecting $\{A, AB\}$ could be due to a decline in the covariate indices of goods $\{C,...,J\}$ so that individuals switch to selecting $\{A, AB\}$. This explains another restriction $\Delta_{s,t}\delta_k < 0$ for some $k\in \{C,...,J\}$ in the last part of Proposition \ref{prop:three}.

\subsection{Extension: Cross-Sectional Models}\label{sec:cross}

In the paper, I studied the multinomial choice models with panel data. This section briefly describes how the identification strategy can be applied to cross-sectional multinomial choice models. 

Now consider the model with cross-sectional data. The utility $u_{ij}$ of consumer $i$ from selecting good $j\in \{A, B\}$ depends on the covariate $X_{ij}$ and the error term $\epsilon_{ij}$ in the following form:
\[u_{ij}=X_{ij}'\beta_0+\epsilon_{ij}. \]

The utility of the outside option is normalized to zero: $u_{iO}=0$. The incremental utility $\Gamma_{iAB}$ from selecting bundle $AB$ is given as
\[\Gamma_{iAB}=u_{iAB}-u_{iA}-u_{iB}. \]  

I consider the same assumption as in Assumption \ref{ass:par}, which assumes that the incremental utility only depends on the covariate $Z_i$:
\[\Gamma_{iAB}=\Gamma(Z_i). \]

The covariate $Z_i$ is suppressed, so we use $\Gamma_0$ to represent $\Gamma(Z_i)$.
With cross-sectional data, the stationarity condition in Assumption \ref{ass:sta} needs to be adjusted. Let $\epsilon_i=(\epsilon_{iA}, \epsilon_{iB})$ and $X_i=(X_{iA}, X_{iB})$. I assume that the error term $\epsilon_i$ is independent of $X_i$.
This assumption imposes an independence condition between the error term $\epsilon_i$ and covariate $X_i$ but still allows the error term to be freely correlated across choices.

Under this assumption, a similar identification result as Proposition \ref{prop1} can be derived by exploiting variation in covariates. Given two different values of the covariate $X_i=x$ and $X_i=\tilde{x}$, let $\delta_j=x_j'\beta_0$ and $\tilde{\delta}_j=\tilde{x}_j'\beta_0$. The following conditions hold for any $(x, \tilde{x})$:

\begin{enumerate}

\item comparisons of CCP of choice $j\in \mathcal{C} $: 
\begin{equation*}
\begin{aligned}
P(\{j\}\mid x)>  P_t(\{j\} \mid \tilde{x})  \Longrightarrow   \exists k\neq j \ \text{s.t.} \   \delta_{j}-\tilde{\delta}_j> \delta_{k}-\tilde{\delta}_k;
\end{aligned}
\end{equation*}

\item comparisons of the demand for good $\ell \in \{A, B\}$:
\begin{equation*}
\begin{aligned}
P\big( &\{\ell, AB\}\mid x \big)>P \big(\{\ell, AB\}\mid \tilde{x}\big) \Longrightarrow \\
&\{ \delta_{\ell}-\tilde{\delta}_{\ell} > 0\} \lor \Big \{ (\delta_{\ell}-\tilde{\delta}_{\ell}+ \sign(\Gamma_0)( \delta_{\ell_{-1} }-\tilde{\delta}_{\ell_{-1} })> 0,  |\Gamma_0|> -(\delta_{\ell}-\tilde{\delta}_{\ell}) \Big\},
\end{aligned}
\end{equation*}
where $\ell_{-1} \in \{A, B\} $ and $\ell_{-1}\neq \ell$;

\item comparisons of the sum of CCP of two choices:
\begin{equation*}
\begin{aligned}
 P(\{AB\}\mid  x)&+P_t(\{O\}\mid \tilde{x})> 1 \Longrightarrow \\
 \Big\{ \Gamma_0&>-\min\{ \delta_{A}-\tilde{\delta}_A,  \delta_B-\tilde{\delta}_B \}  \Big\} \land \{ \delta_{A}-\tilde{\delta}_A+\delta_B-\tilde{\delta}_B  >0\}; \\
P(\{A\}\mid x)& +P(\{B\}\mid  \tilde{x})> 1 \Longrightarrow  \\ 
\Big\{\Gamma_0&<\min \{\delta_{A}-\tilde{\delta}_A, -(\delta_{B}-\tilde{\delta}_B)\} \Big\} \land \{  \delta_{A}-\tilde{\delta}_A- (\delta_{B}-\tilde{\delta}_B) > 0 \}. \\
 \end{aligned}
 \end{equation*} 
 
\end{enumerate}

\subsection{Extension: Nonseparable Utility Functions}\label{subsec:nonsep}

The paper focused on an additive and separable utility function which is commonly used in the literature on discrete choice models. 
This section studies a more general class of utility functions that can be nonseparable between observed covariates and unobserved heterogeneity. This class of utility functions can allow flexible interactions between observed covariates, fixed effects, and error terms.

Consider that there are two goods $\{A, B\}$. 
The utility $u_{ijt}$ for consumer $i$ from selecting good $j\in \{A, B\}$ at time $t$  still depends on the three crucial components: covariate $X_{ijt}$, unobserved agent-level fixed effects $\alpha_{ij}$, and unobserved error terms $\epsilon_{ijt}$. Different from Model \eqref{model:u1}, this section studies a more general utility function as follows: for $j\in \{A, B\}$,
\begin{equation}\label{model:u2}
 u_{ijt}=u(X_{ijt}'\beta_0, \alpha_{ij}, \epsilon_{ijt}).
 \end{equation}
 
The function $u$ can be potentially unknown to researchers and could be nonseparable between the three components. The only restriction on the function $u$ is a monotonicity condition stated in Assumption \ref{ass:mon}. This class of utility functions can admit richer preferences of consumers. 
%the covariate index $X_{ijt}'\beta_0$, the fixed effects $\alpha_{ij}$, and the error term $\epsilon_{ijt}$.

The utility of the outside option $O$ is normalized as zero: 
\[u_{iOt}=0.\]

We are still interested in the incremental utility $\Gamma_{it}$ from consuming bundle $AB$ compared to choosing a single good:
\[\Gamma_{it}=u_{iABt}-u_{iAt}-u_{iBt}. \]

I consider the same assumption for the complementarity which can only depend on observed covariate $Z_i$ as in Assumption \ref{ass:par}. The covariate $Z_i$ is suppressed so the complementarity $\Gamma_0$ is written as a constant.
%Similar to model \eqref{model:u1}, the utility for consumer $i$ from selecting choice  $j$  at time $t$  is specified as follows
%\begin{equation}\label{model:u2}
%\begin{aligned}
%&u_{iAt}=u(X_{iAt}'\beta_0, \alpha_{iA}, \epsilon_{iAt}), \\
%&u_{iBt}=u(X_{iBt}'\beta_0, \alpha_{iB}, \epsilon_{iBt}), \\
%&u_{iABt}=u_{iAt}+u_{iBt}+\Gamma_{it},\\%(Z_i, \gamma_0) \\
%&u_{iOt}=0,
%\end{aligned}
%\end{equation}

The utility function in \eqref{model:u2} has also been studied by \cite{gao2020}, while their paper focuses on the identification of the coefficient $\beta_0$.  My paper allows for purchasing bundle $AB$ and provides identification results for the complementarity $\Gamma(z)$ between the two goods. Following \cite{gao2020},  I assume a monotonicity condition on the utility function to the covariate index $X_{ijt}'\beta_0$.

\begin{ass} (Monotonicity)\label{ass:mon}
The utility $u(\delta, \alpha, \epsilon)$ is weakly increasing in the index $\delta$ for every realization $(\alpha, \epsilon)$, i.e.
\begin{equation*}
\text{for any } \ (\alpha, \epsilon),  \quad u(\tilde{\delta}, \alpha, \epsilon)\geq u(\delta, \alpha, \epsilon) \quad \text{if} \ \tilde{\delta}\geq \delta.
\end{equation*}
\end{ass}

Assumption \ref{ass:mon} only requires monotonicity with respect to the covariate index but imposes no restriction on unobserved fixed effects and error terms. 
The additively separable utility function in Equation \eqref{model:u1} is nested in Assumption \ref{ass:mon}. The nonseparable utility function not only admits flexible interactions between observed characteristics and unobserved heterogeneity but also allows for nonlinear functions of the covariate $X_{ijt}$ such as exponential functions or higher-order polynomial functions. 

%The following example introduces a nonseparable utility function satisfying assumption \ref{ass:mon}:
%\begin{example}[Exponential Utility]
%\begin{equation*}
%u(\delta, \alpha, \epsilon)=\left\{ 
% \begin{aligned}
%\frac{1-e^{-\delta \alpha \epsilon}}{\alpha \epsilon}    \qquad  \text{if} \ \ & \alpha\epsilon\neq 0 \\
%\delta    \qquad  \qquad  \text{if}  \ \ & \alpha\epsilon= 0 \\
%\end{aligned}
% \right.
%\end{equation*}
%\end{example}
%
%\vspace{0.2cm}
%This utility function form is similar to the CRRA function, and $\alpha \epsilon$ captures the risk aversion of the consumer. It allows for interactions between the observed covariate index $\delta$ and unobserved heterogeneity $(\alpha, \epsilon)$, therefore the marginal utility from covariates can depend on unobserved heterogeneity $\alpha\epsilon$. This class of utility functions can accommodate heterogeneous preferences which can not be admitted in the additive and separable utility function. 
%There are also  other examples of nonseparable utility functions satisfying the monotonicity assumption, such as higher-order polynomial functions.  

Recall that $D_{\ell}=\{\ell, AB\}$ for $\ell \in \{A, B\}$. The next proposition characterizes identifying restrictions for the parameter $\theta_0=(\beta_0, \Gamma_0)$. 

\begin{prop} \label{prop:nonsep}
Under Assumptions \ref{ass:par}-\ref{ass:sta} $\&$ \ref{ass:mon}, the following conditions for any $(x_{s}, x_{t})$ and $ s\neq t\leq T$,
 
\begin{enumerate}
\item[{(1)}]  comparisons of CCP of choice $j\in \mathcal{C}$,
% When $j\in\{A, B\}$, let $j_{-1}\neq j$ denote the other goods in $\{A, B\}$.
%\begin{equation*}
%\Delta X_{ij}'\beta_0 \geq 0, \ \Delta X_{ij_{-1}}'\beta_0\leq 0 \  \Longrightarrow \Pr(Y_{it_1}=j|W_{ist}) \geq \Pr(Y_{it_2}=j|W_{ist})
%\end{equation*}
\begin{equation*} \label{gen1}
\begin{aligned} 
 P_s(\{j\}\mid x_{s}, x_{t}) > & P_t(\{j\} \mid x_{s}, x_{t}) \Longrightarrow  \\
\big\{& (-1)^{\mathbbm{1}[j\in D_A ]}\Delta_{s,t} \delta_{A} < 0 \big\} \lor \big\{ (-1)^{\mathbbm{1}[j\in D_B] }\Delta_{s,t} \delta_{B}<0 \big\} ;
\end{aligned}
\end{equation*}

\item[{(2)}] comparisons of the demand for good $\ell \in \{A, B\}$:
\begin{equation*}\label{gen2}
\begin{aligned}
P_s(D_{\ell}\mid x_{s}, x_{t})> P_t(D_{\ell} \mid & x_{s}, x_{t}) \Longrightarrow \\
& \{ \Delta_{s, t} \delta_{\ell}> 0\} \lor \big\{ \sign(\Gamma_0 )   \Delta_{s, t} \delta_{\ell_{-1}}>  0   \big\},
  \\
\end{aligned}
\end{equation*}
where $\ell_{-1}\in \{A, B\} $ and $\ell_{-1}\neq \ell$.

\end{enumerate}

\end{prop}

Similar to Proposition \ref{prop1}, Proposition \ref{prop:nonsep} derives identifying conditions for the parameter $\theta_0$ using intertemporal variation in conditional choice probabilities over any two periods. The main difference is that the additively separable utility function in \eqref{model:u1} imposes restrictions on both the direction and the degree of how covariate indices affect agents' utility $u_{ijt}$, while the nonseparable function in \eqref{model:u2} only assumes a monotonicity condition but is flexible about the degree of how covariate indices affect the utility. Therefore, the identified set in Proposition \ref{prop:nonsep}  is wider compared to the result in Proposition \ref{prop1}, while the results in Proposition \ref{prop:nonsep} are robust to the specifications of utility functions. The proof of Proposition  \ref{prop:nonsep} is similar to Appendix \ref{proof:prop1} so it is omitted here.

In Proposition \ref{prop:nonsep},  the first condition can identify the coefficient $\beta_0$ by using the conditional  probability of a single choice. The intuition is as follows: if the conditional probability of selecting choice $j$ increases, then it must be that either good $j$ becomes better (covariate index increases) or other goods become worse (covariate index decreases). 

The second condition in Proposition \ref{prop:nonsep} identifies the sign of  $\Gamma_0$ using variation in conditional demand over time. Similar to condition \eqref{id2} in Proposition \ref{prop1}, the idea is to exploit the relationship between covariate indices and demand under different complementarity relationships. When two goods are complements, an increase in the covariate index of good $A$ will drive up the demand for good $B$ since people will switch to choosing the bundle $AB$. Then, if a decline in the demand for good $B$ is observed, it can be inferred that two goods are substitutes $\Gamma_0 \leq 0$. We are unable to bound the value of the complementarity since the effect of the covariate index on the utility is unknown with this nonseparable utility function.

\subsection{Discussion on the Utility Specification} \label{sec:dis}
%The uitilit specification in this paper links the utility of $AB$ with the utility of individual goods 

In this section, we discuss the connection between the utility specification in this paper and the conventional discrete choice model. In particular, take a three-good version of \cite{pakes2019}'s model:
\begin{equation}\label{model:stan}
\begin{aligned}
&u_{iAt}=X_{iAt}'\beta_0+\alpha_{iA}+\epsilon_{iA}, \\
&u_{iBt}=X_{iBt}'\beta_0+\alpha_{iB}+\epsilon_{iB}, \\
&u_{iCt}=X_{iCt}'\beta_0+\alpha_{iC}+\epsilon_{iCt},\\
&u_{iOt}=0.
\end{aligned}
\end{equation}
One could rename $C=AB$ and interpret the bundle $AB$ as an additional good.

%The utility function in Model \eqref{model:u1} directly specifies the utility of $AB$ by linking it with the utility of individual goods, given by

In my model, I instead directly specify the utility of $AB$ by linking it with the utility of individual goods, given by
\begin{equation} \label{model:com}
\begin{aligned}
&u_{iAt}=X_{iAt}'\beta_0+\alpha_{iA}+\epsilon_{iA}, \\
&u_{iBt}=X_{iBt}'\beta_0+\alpha_{iB}+\epsilon_{iB}, \\
&u_{iABt}=u_{iAt}+u_{iBt}+\Gamma(Z_i),\\
&u_{iOt}=0.
\end{aligned}
\end{equation}

The above specification assumes that the incremental utility from selecting both goods compared to the sum of utilities of individual goods only depends on observed characteristic $Z_i$. This specification can be viewed as imposing the following condition on \eqref{model:stan}:
%It essentially imposes the following condition on \eqref{model:stan}:
%Compared to treating $AB$ as a separate good in \eqref{model:stan}, the utility specification in \eqref{model:com} can be viewed as imposing the following condition:
\begin{equation}\label{equ:com}
X_{iABt}'\beta_0+\alpha_{iAB}+\epsilon_{iABt}=(X_{iAt}+X_{iBt})'\beta_0+\alpha_{iA}+\alpha_{iB}+\epsilon_{iAt}+\epsilon_{iBt}+\Gamma(Z_i).
\end{equation}
Notice that the complementarity $\Gamma(Z_i)$ from consuming both goods is assumed to only depend on observed characteristics, but could be sourced from any combination of the covariate index, the fixed effects, and the unobserved shock.\footnote{In this model, we can only identify the total complementarity arising from all three components since dividing the complementarity term arbitrarily into the three parts will not affect consumers' choices.} For example, suppose $\epsilon_{ijt}$ represents the unobserved quality of the product. The quality contribution to the utility of the bundle,  $\epsilon_{iABt}$, might not equal $\epsilon_{iAt}+\epsilon_{iBt}$, the sum of the quality contributions to $u_A$ and to $u_B$. This accounts for substitution/complementarity coming from the unobserved quality. However, the difference, $\epsilon_{iABt}-\epsilon_{iAt}-\epsilon_{iBt}$, is assumed to only depend on observables. This structure enables us to test and identify the substitution and complementarity relationship between goods.

Condition \eqref{equ:com}
 assumes that the complementarity only depends on observed characteristics. A more general specification is to incorporate unobserved heterogeneity in the complmentarity, given by
\[X_{iABt}'\beta_0+\alpha_{iAB}+\epsilon_{iABt}=(X_{iAt}+X_{iBt})'\beta_0+\alpha_{iA}+\alpha_{iB}+\epsilon_{iAt}+\epsilon_{iBt}+\Gamma_{it}, \]
where $\Gamma_{it}$ is an unobserved individual-specific error term. Thus, it allows for heterogeneous complementarity relationships across individuals that are not captured by observed characteristics. Section \ref{sec:exte} explores this specification but only weaker identification results can be obtained.

\subsection{Testing Substitutability} \label{sec:test}

In certain applications, we may think that two goods are substitutes  and want to test the substitutability relationship between goods. 
The idea of testing the substitutability is similar to the analysis in Section \ref{subsec:test}. Testing the substitutability is equivalent to testing $\Gamma_0\leq 0$ (covariate $Z_i$ is suppressed). 
Then, the null hypothesis $H'_0$ and alternative hypothesis $H'_1$ are given as follows:
\begin{equation*} 
H'_0: \Gamma_0\leq 0 \qquad H'_1: \Gamma_0>0.
\end{equation*}

The main idea is still to construct moment inequalities that only depend on observed variables under the null hypothesis $H'_0$. To construct moment restrictions, we will exploit the relationship between the demand and covariate indices of the two goods under the substitution relationship between goods. The intuition is described as follows. If the two goods are substitutes (under $H'_0$), an improvement in good $A$ and a decline in good $B$ should result in an increase in demand for good $A$ and a decrease in demand for good $B$. If instead a decrease in demand for good $A$ or an increase in demand for good $B$ is observed, it can be deduced that the two goods are complements and the null hypothesis $H'_0$ is not supported.

To characterize testable implications, the first step is to infer the sign of covariate indices of the two goods such that good $A$ becomes better and good $B$ becomes worse. Let $\xi_{s, t}^{2}(x_{s}, x_{t})$ be an indicator for increasing probabilities of all choices $\{A, AB, O\}$, defined as
\[\xi_{s, t}^{2}(x_{s}, x_{t})= \mathbbm{1}\big\{P_s(\{j\} \mid x_{s}, x_{t})- P_t(\{j\} \mid x_{s}, x_{t})\geq 0,  \ \forall j \in \{A, AB, O\}  \big\}. \]

When an increase in conditional probabilities of all choices $j\in \{A, AB, O\}$ is observed, we can infer that
\begin{equation*}
\xi_{s, t}^{2}(x_{s}, x_{t})=1 \Longrightarrow \Delta_{s,t}\delta_{A} \geq 0, \  \Delta_{s,t}\delta_{B} \leq 0.
\end{equation*}

Since there is an increase in probabilities of all choices except good $B$,  it can only be deduced that the covariate index of good $A$ increases and that of good $B$ decreases.
Given the above sign of covariate indices of the two goods, the conditional demand for good $A$ should increase and the demand for good $B$ should decrease under the null hypothesis of the two goods being substitutes. When an increase in demand for good $A$ or a decline in demand for good $B$ is observed, we can conclude that two goods are complements and reject the null hypothesis. The following proposition characterizes the testable implications of the null hypothesis $H'_0$.

\begin{prop}\label{prop:test2}
Under Assumptions \ref{ass:par}-\ref{ass:sta}, the following conditional moment inequalities hold under the null hypothesis $H'_0$: 
\begin{equation*}    
\left\{
\begin{aligned}
&E\Big[ \xi_{s, t}^{2}(x_{s}, x_{t})\big(\mathbbm{1}\{Y_{is}\in D_{A}\}- \mathbbm{1}\{Y_{it}\in D_{A} \} \big)  \bigm| x_{s}, x_{t} \Big] \geq 0; \\
&E\Big[  \xi_{s, t}^{2}(x_{s}, x_{t}) \big(\mathbbm{1}\{Y_{is}\in D_{B}\}- \mathbbm{1}\{Y_{it}\in D_{B} \} \big)  \bigm| x_{s}, x_{t} \Big] \leq 0,
\end{aligned}
\right.
\end{equation*}
for any $(x_s, x_t)$ and any $s\neq t \leq T$.
\end{prop}

%Proposition \ref{prop:test2} exploits variation in covariates indices and demand for the two goods to characterize conditional moment inequalities under the null hypothesis. The moment inequalities are free from fixed effects and error terms, and only depend on observed variables $(X_{it}, Y_{it})$. Therefore, we can directly test these restrictions using methods in the literature.

\subsection{More Simulation Results} \label{sec:sim_more}

\subsubsection{Longer Panels $T\geq 2$}

This section studies the performance of the point estimator but considers longer panels $T=\{2, 3, 4\}$. I compare the two-step estimator with the parametric estimator as well as two other estimators for $\beta_0$ described in the paper. I study the same simulation setup for covariates and the same four DGP designs for fixed effects and error terms. The sample size is $N=1000$ and the simulation repetition is $B=500$.

Table \ref{table:gammalong} and Table \ref{table:betalong} compare the performance of the two-step estimator with other estimators for the complementarity parameter $\gamma_0$ and the utility coefficient $\beta_0$ under longer panels. The performance of the two-step estimator improves when the length of panels increases and the performance is robust in all designs. The parametric estimator performs better with longer panels, but it still can have a large bias when there is any misspecification (designs 2-4). 
The two-step estimator outperforms other estimators in designs 2-4 regardless of the length of periods and keeps the advantage of performing more robustly.

%and other estimators perform better with longer panels if the assumptions are correctly specified (design 1), but can become worse if there is any misspecification (designs 2-4). 
 
\begin{table}[!htbp]
\centering
\caption{Performance Comparisons for $\hat{\gamma}$ (longer panel)}
\label{table:gammalong}
\begin{tabular}{cc |cccc|cccc}
\hline
\hline
 \multirow{2}{*}{$T$}&\multirow{2}{*}{Design}&\multicolumn{4}{c|}{Two-Step Est}  &  \multicolumn{4}{c}{Parametric Est}  \\
\cline{3-10} 
&  & Err & SD & rMSE & MAD  & Err & SD  & rMSE & MAD \\
\hline
\multirow{5}{*}{$T=2$} & design 1  &0.025 & 0.263  &  0.315 & 0.267  & 0.007&  0.101 &0.101  &0.079  \\ [1ex]       
& design 2  &0.026 & 0.309 &  0.341 & 0.278  & 0.058 &  0.344 &0.741 &0.663  \\ [1ex]                                                                                                                                                                                                                                                                                                                                                                                                                                                                                                                                                                                                                                                                   
&design 3  & 0.023 & 0.287 &  0.306 &  0.251 &0.083 & 0.288 & 1.019 &0.977  \\[1ex]   
&design 4  & 0.025 &0.329 & 0.332 & 0.264 & 0.126 & 0.681 & 1.764 &1.628 \\[1ex]  
\hline
\multirow{5}{*}{$T=3$} & design 1  & 0.018&  0.204 &0.233 & 0.187   &0.006  &  0.079 &  0.079 & 0.063 \\ [1ex]  
& design 2  &  0.022   &  0.291  &  0.293 &0.238   & 0.063  &  0.311  & 0.782 &  0.721    \\ [1ex]  
& design 3  & 0.018   & 0.233  &0.233 & 0.187    & 0.056 &  0.232 &  0.674  &  0.634    \\ [1ex]  
& design 4  &  0.021 &  0.284 &  0.286  & 0.229 &   0.115 & 0.527   & 1.533  & 1.440     \\ [1ex]  
\hline
\multirow{5}{*}{$T=4$} & design 1  & 0.018   & 0.181 & 0.231 &  0.193 &  0.006  & 0.075  &  0.075   &  0.059  \\  [1ex]       
& design 2  &   0.021   &   0.287 &   0.290  &  0.230  &   0.066   & 0.282   &  0.796 & 0.746  \\  [1ex]                                                                                                                                                                                                                                                                                                                                                                                                                                                                                                                                                                                                                                                                    
&design 3  & 0.015   &   0.194 &   0.200  &   0.162   &  0.041  &  0.179  &   0.487  & 0.453 \\  [1ex]  
&design 4  &  0.017   & 0.229   &  0.229 &  0.181   & 0.108 & 0.385 & 1.379 &   1.324    \\ [1ex]   
\hline
\end{tabular}
\end{table}

\begin{table}[!htbp]
\centering
\caption{Performance Comparisons for $\hat{\beta}$ (longer panel)}
 \label{table:betalong}
\centering
\begin{tabular}{cc |cc|cc|cc|cc}
\hline
\hline
& & \multicolumn{4}{c|}{Estimators with bundles} &  \multicolumn{4}{c}{Estimators assuming no bundles}\\
\hline
 \multirow{2}{*}{$T$}&\multirow{2}{*}{Design}&\multicolumn{2}{c|}{Two-Step Est.} &  \multicolumn{2}{c|}{Parametric Est.}  & \multicolumn{2}{c|}{FE Logit Est.}  & \multicolumn{2}{c}{Semi. Est.} \\
\cline{3-10}
&  & rMSE & MAD & rMSE & MAD  & rMSE & MAD  & rMSE & MAD \\
\hline
\multirow{5}{*}{$T=2$} & design 1  &  0.126 & 0.101 & 0.072& 0.056  & 0.165 &  0.130 &0.155 & 0.131   \\[1ex]  
&design 2  &  0.125 & 0.101 &  0.432 & 0.389  &  0.243  & 0.187 & 0.161 & 0.134 \\[1ex]  
&design 3  &  0.128 & 0.100 &  0.304 & 0.279  &  0.168 & 0.131 &  0.149 &0.124 \\[1ex]  
&design 4 & 0.121 & 0.095 & 0.647 & 0.593 & 0.256 & 0.200 & 0.171 & 0.143 \\[1ex]  
\hline
\multirow{5}{*}{$T=3$} & design 1  & 0.087 & 0.069 &0.052 &0.042 &0.107 &0.084 &0.130 &0.108  \\  [1ex]  
&design 2  & 0.083 & 0.064 & 0.385 & 0.360 & 0.178 &0.149 &0.123 &0.103 \\ [1ex]  
&design 3  & 0.090 & 0.070 & 0.214 &0.198 &0.106 &0.084 &0.124 &0.102 \\ [1ex]  
&design 4  & 0.085  &  0.069 & 0.539 & 0.509 &0.189 & 0.157 & 0.131 &0.109  \\ [1ex]  
\hline
\multirow{5}{*}{$T=4$} & design 1  & 0.081  & 0.066  & 0.045  & 0.036  & 0.079  & 0.062 &0.117   & 0.095   \\  [1ex]  
&design 2  & 0.069   &  0.050  & 0.081  &  0.058   &   0.145  & 0.125  &  0.095  &  0.060  \\ [1ex]  
&design 3  &0.075 & 0.060  &  0.165&  0.153 & 0.080   & 0.064 &0.116 &0.096  \\ [1ex]  
&design 4  &  0.074 & 0.059  &  0.488&  0.465 & 0.163  & 0.138 &0.119 &0.097   \\ [1ex]  
\hline
\end{tabular}
\end{table}

\newpage

\subsubsection{Partial Identification}\label{sec:sim_par}

This section examines the performance of set estimators when covariates have bounded support. The parametric approach is also implemented for comparison, which achieves point identification regardless of the support of covariates. 

The simulation setup is similar except the support of some covariates is bounded. Let $d_x=2$ and $d_z=2$ denote the dimension of the covariates $X_{i\ell t}$ and $Z_i$ respectively, for $\ell \in \{A, B\}$. 
In each simulation,  $X_{iA t}$ is drawn from the uniform distribution $U[-3, 3]$ and $X_{iBt}$ is drawn from the normal distribution $\mathcal{N}(0, 2)$. The covariates are independent across choices $\ell\in \{A, B\}$ and time $t\leq T$.  Let the first element of $Z_i$ be drawn from the uniform distribution $U[0, 4]$, and the second element is drawn from $U[-2, 2]$. 
The true parameters are set as $\beta_0=\gamma_0=(1,1)$.  We study the same four designs of fixed effects and error terms as before. 
Consider that the sample size is $N\in \{1000, 4000\}$, the length of periods is $T=2$, and the repetition number is $B=500$.

The parameter is only partially identified since the support of $X_{iAt}$ and $Z_i$ is bounded. Following  \cite*{chernozhukov2007}, the set estimator for the identified set $\Theta_I$ is proposed as
\begin{equation*}
\hat{\Theta}_{\hat{c}_N}=\Big\{\theta\in \Theta: \hat{\Omega}_N(\theta) \leq \inf_{\theta\in \Theta} \hat{\Omega}_N(\theta)+\hat{c}_N/a_N \Big\},
\end{equation*}
where $a_N$ denotes the uniform convergence rate of the sample objective function $ \hat{\Omega}_N$, which is $a_N \approx N^{1/4}$ when the neural network estimator is used in the first step; $\hat{c}_N$ is chosen as $10^{-4}\log(N)$ so it satisfies $\hat{c}_N/a_N \rightarrow 0$. I divide the interval $[-5, 5]$ into $k=100$ grids and apply grid search to find the identified set. 

Let $(\hat{\beta}_l, \hat{\beta}_u)$ and $(\hat{\gamma}_l, \hat{\gamma}_u)$ denote the lower bound and upper bound for $\beta_0$ and $\gamma_0$, respectively. Let $\hat{\beta}_{par}, \hat{\gamma}_{par}$ denote the parametric estimator for $\beta_0$ and $\gamma_0$. To compare the two approaches, I still report the standard deviation (SD),  root mean-squared error (rMSE), and mean of absolute deviation (MAD).

Table \ref{table:gamma_par} and \ref{table:beta_par} compare the performance of the lower/upper bounds in this paper with the parametric approach. The comparison between the two approaches keeps similar patterns as in the point identification section. Although the performance of the set estimator is slightly worse compared to the point estimator with large support, it still performs more robustly than the parametric estimator concerning specifications of fixed effects or distributional assumptions of error terms. The parametric approach has a smaller bias under correct specification (design 1), but can be severely biased with misspecifications (design 2-4). The set estimator proposed in this paper performs robustly with different specifications of fixed effects and error terms.

\begin{table}[!htbp]
\centering
\caption{Performance Comparisons for $\hat{\gamma}$}
\label{table:gamma_par}
\begin{tabular}{cc|ccc|ccc|ccc}
\hline
\hline
 \multirow{2}{*}{$N$}&\multirow{2}{*}{Design}&\multicolumn{6}{c|}{Two-Step Est.}  &  \multicolumn{3}{c}{Parametric Est.}  \\
\cline{3-11} 
& & \multicolumn{3}{c|}{ $\hat{\gamma}_l$ } &  \multicolumn{3}{c|}{ $\hat{\gamma}_u$  } &  \multicolumn{3}{c}{$\hat{\gamma}_{par}$}  \\
\cline{3-11} 
&  & SD & rMSE & MAD  & SD & rMSE & MAD & SD & rMSE & MAD \\
\hline
\multirow{5}{*}{1000} & design 1  & 0.309 & 0.353 &   0.285 & 0.318 & 0.483 & 0.410  & 0.083 & 0.083 & 0.067\\ [1ex]       
& design 2  & 0.365 &0.367 &0.292 &0.397 &0.492 &0.394 & 0.389  & 0.946 & 0.864  \\ [1ex]                                                                                                                                                                                                                                                                                                                                                                                                                                                                                                                                                                                                                                                                   
&design 3  & 0.376 &0.510 &0.418 & 0.406 & 0.739 & 0.635 & 0.241 &1.061 & 1.032  \\[1ex]   
&design 4  & 0.408 & 0.538 & 0.419  & 0.456 & 0.766 & 0.634  & 0.690 & 2.670 &   2.580  \\[1ex]  
\hline
\multirow{5}{*}{4000} & design 1  & 0.205 & 0.208 &0.170  & 0.211 & 0.334 &0.283  & 0.043 & 0.043 & 0.034   \\ [1ex]      
& design 2  & 0.221 & 0.239 & 0.189 & 0.231 & 0.459 & 0.403 & 0.197 & 0.745 &0.718  \\ [1ex]                                                                                                                                                                                                                                                                                                                                                                                                                                                                                                                                                                                                                                                                    
&design 3  & 0.212 & 0.242 & 0.195 & 0.233 & 0.538 & 0.488 & 0.149 & 1.047 & 1.036 \\[1ex]    
&design 4  & 0.204 & 0.290 & 0.238 & 0.232 &0.588 & 0.543 & 0.409 & 2.767 & 2.734   \\[1ex]   
\hline
\end{tabular}
\end{table}

\begin{table}[!htbp]
\centering
\caption{Performance Comparisons for $\hat{\beta}$}
\label{table:beta_par}
\begin{tabular}{cc|ccc|ccc|ccc}
\hline
\hline
 \multirow{2}{*}{$N$}&\multirow{2}{*}{Design}&\multicolumn{6}{c|}{Two-Step Est.}  &  \multicolumn{3}{c}{Parametric Est.}  \\
\cline{3-11} 
& & \multicolumn{3}{c|}{ $\hat{\beta}_l$ } &  \multicolumn{3}{c|}{ $\hat{\beta}_u$ } & \multicolumn{3}{c}{ $\hat{\beta}_{par}$ }  \\
\cline{3-11} 
&  & SD & rMSE & MAD  & SD & rMSE & MAD & SD & rMSE & MAD \\
\hline
\multirow{5}{*}{1000} & design 1  & 0.121 &0.129 & 0.107 & 0.132 & 0.132 &0.107 & 0.068 & 0.069 & 0.054  \\ [1ex]       
& design 2  & 0.125 & 0.136 & 0.113 & 0.131 &0.131 & 0.106 &0.160 &0.186 &0.148 \\ [1ex]                                                                                                                                                                                                                                                                                                                                                                                                                                                                                                                                                                                                                                                                   
&design 3  & 0.125 &0.129 & 0.104  & 0.133 & 0.138 & 0.110 & 0.138 &0.460 & 0.438   \\[1ex]   
&design 4  & 0.127 & 0.131 &0.109 & 0.134 & 0.136  & 0.109 & 0.253 & 0.627 & 0.573 \\[1ex]  
\hline
\multirow{5}{*}{4000} & design 1  &0.071  & 0.090 &0.077  & 0.079 &0.081 &0.069  & 0.032 &0.032 &0.026    \\ [1ex]      
& design 2  & 0.074 & 0.094 & 0.079 & 0.077 & 0.077 & 0.066 & 0.085 &0.092 & 0.070 \\ [1ex]                                                                                                                                                                                                                                                                                                                                                                                                                                                                                                                                                                                                                                                                    
&design 3  & 0.069 & 0.079 & 0.067 & 0.076 & 0.086 & 0.073 &0.069 &0.424 & 0.419\\[1ex]    
&design 4  & 0.072 &0.084 &0.071 &0.079 & 0.086 &0.072 & 0.132 & 0.596 & 0.581 \\[1ex]   
\hline
\end{tabular}
\end{table}

\subsection{Proof of Proposition \ref{hetero}} \label{proof:hetero}
\begin{proof}

The conditional demand for one good can be expressed as a mixture of two groups: one group is people to whom the two goods are complements $\Gamma_{it}\geq 0$ and the other is people to whom the two goods are substitutes $\Gamma_{it}<0$. Therefore, the demand for good $A$ (or $B$) conditional on the covariate and the fixed effects is given as follows:
\begin{equation*}
\begin{aligned}
\Pr(Y_{it}\in D_A\mid \alpha_i, x_s, x_t)=\Pr(&Y_{it}\in D_A\mid \alpha_i, x_s,  x_t, \Gamma_{it}\geq 0)\Pr(\Gamma_{it}\geq 0\mid \alpha_i)+\\
&\Pr(Y_{it}\in D_A\mid \alpha_i, x_s, x_t, \Gamma_{it}< 0)[1-\Pr(\Gamma_{it}\geq 0\mid \alpha_i)].
\end{aligned}
\end{equation*}
The above equation holds since Assumption \ref{ass:gam} (conditional independence) implies $\Pr(\Gamma_{it}\geq 0\mid \alpha_i, x_s, x_t)=\Pr(\Gamma_{it}\geq 0\mid \alpha_i)$. 
% \ref{ass:gam}

The main strategy is to use intertemporal variation in the conditional demand for good $A$ to bound the probability $\Pr(\Gamma_{it}\geq 0\mid \alpha_i)$. It can be shown that when $(x_s, x_t)\in \mathcal{X}_{s,t}^{1}$, the conditional demand $\Pr(Y_{it}\in D_A\mid \alpha_i, x_s,  x_t, \Gamma_{it}\geq 0)$ increases at time $s$ compared to time $t$, which will be proven later. The variation in the conditional demand given $\Gamma_{it}<0$ can be bounded as $[-1, 1]$ since it is the difference between two probabilities. Therefore, the variation in the aggregate demand for good $A$ can be bounded as: for any $(x_s, x_t)\in \mathcal{X}_{s,t}^{1}$,
\begin{equation*}
\Pr(Y_{is}\in D_A\mid \alpha_i, x_s, x_t)-\Pr(Y_{it}\in D_A\mid \alpha_i, x_s, x_t)\geq 0+(-1)*[1-\Pr(\Gamma_{it}\geq 0\mid \alpha_i)].
\end{equation*}

By taking expectation over the fixed effect $\alpha_i$ conditional on the covariate, the probability $\eta=\Pr(\Gamma_{it}\geq 0)$ can be bounded above as follows: for any $(x_s, x_t)\in \mathcal{X}_{s,t}^{1}$,
\begin{equation*}
\eta \leq P_s(D_A\mid x_s, x_t)-P_t(D_A\mid  x_s, x_t) +1.
\end{equation*}

Since the probability $\eta$ does not depend on covariates and it is stationarity over time under Assumption \ref{ass:sta_gam}, it can be bounded by taking the infimum over all values of the covariates and any two periods. Moreover, the variation in the demand for good $B$ can also be exploited to bound the probability $\eta$ similarly. Therefore, the upper bound for $\eta$ can be established as follows:
\begin{equation*}
\eta \leq \inf_{(x_s, x_t)\in \mathcal{X}_{s,t}^{1}, \ell\in\{A, B\}, (s,t)\leq T}\Big\{ P_s(D_{\ell}\mid x_s, x_t)-P_t(D_{\ell}\mid x_s, x_t) \Big\}+1=U_{\eta}.
\end{equation*}

Now I need to show that the conditional probability $\Pr(Y_{it}\in D_A\mid \alpha_i, x_s, x_t, \Gamma_{it}\geq 0)$ increases at time $s$ compared to time $t$ when the covariate satisfies $(x_s, x_t)\in \mathcal{X}_{s,t}^{1}$. Let $v_{it}=\epsilon_{it}+\alpha_i$, and let $\mathcal{V}_{D_A}^{\Gamma}(x_t)$ denote the collection of $(v, \Gamma\geq 0)$ such that either choice $A$ or $AB$ is chosen:
\begin{equation*}
\begin{aligned}
\mathcal{V}_{D_A}^{\Gamma}(x_t)=&\{ (v, \Gamma\geq 0)\mid \delta_{At}+v_A\geq \delta_{Bt}+v_B, \   \delta_{At}+v_A\geq 0   \}\equiv \mathcal{V}^{\Gamma}_{1}(x_t), \\
 \cup &\{(v, \Gamma\geq 0)\mid \delta_{At}+v_A+\Gamma\geq 0, \  \delta_{At}+v_A+\delta_{Bt}+v_B+\Gamma \geq 0 \}\equiv \mathcal{V}^{\Gamma}_{2}(x_t).
\end{aligned}
\end{equation*}

The conditional demand $\Pr(Y_{it}\in D_A\mid \alpha_i, x_s, x_t, \Gamma_{it}\geq 0)$ can be expressed as the conditional probability of the set $\mathcal{V}_{D_A}^{\Gamma}(x_t)$:
\begin{equation*}
\Pr(Y_{it}\in D_A\mid \alpha_i, x_s, x_t,\Gamma_{it}\geq 0)=\Pr( (v_{it}, \Gamma_{it})\in \mathcal{V}_{D_A}^{\Gamma}(x_t)\mid \alpha_i, x_s, x_t, \Gamma_{it}\geq 0 ).
\end{equation*}

%\ref{ass:sta_gam}
Assumption \ref{ass:sta_gam} implies that the distribution $(v_{it}, \Gamma_{it})$ conditional on $(\alpha_i, X_{is}, X_{it},\Gamma_{it})$ is stationarity over time. Then I only need to show $\mathcal{V}_{D_A}^{\Gamma}(x_t) \subseteq \mathcal{V}_{D_A}^{\Gamma}(x_{s})$ when the covariate satisfies $(x_s, x_t)\in \mathcal{X}_{s,t}^{1}$, which has the following implication:
\begin{multline*}
\mathcal{V}_{D_A}^{\Gamma}(x_t)\subseteq \mathcal{V}_{D_A}^{\Gamma}(x_s) \Longrightarrow  \\\Pr(Y_{it}\in D_A\mid \alpha_i, x_s, x_t,\Gamma_{it}\geq 0)\leq \Pr(Y_{is}\in D_A\mid \alpha_i, x_s, x_t,\Gamma_{it}\geq 0).
\end{multline*}

To prove $\mathcal{V}_{D_A}^{\Gamma}(x_t) \subseteq \mathcal{V}_{D_A}^{\Gamma}(x_s)$, I will show that for any element $(v, \Gamma)\in \mathcal{V}_{D_A}^{\Gamma}(x_t)$, it also satisfies  $(v, \Gamma)\in \mathcal{V}_{D_A}^{\Gamma}(x_s)$ when $(x_s, x_t)\in \mathcal{X}_{s,t}^{1}$. As shown before, any $(x_s, x_t)\in \mathcal{X}_{s,t}^{1}$ satisfies $\delta_{At}\geq0, \delta_{Bt}\geq 0$. 
I discuss two cases to prove the statement:  $(v, \Gamma)\in \mathcal{V}_{1}^{\Gamma}(x_t)$ and $(v, \Gamma)\in \mathcal{V}_{2}^{\Gamma}(x_t)$. 

Case 1: $(v, \Gamma)\in \mathcal{V}_{2}^{\Gamma}(x_t)$. According to the definition of the set $\mathcal{V}_{2}^{\Gamma}(x_t)$, it increases with respect to the covariate index for good $A$ and good $B$. Therefore, we know that $(v, \Gamma) \in \mathcal{V}_{2}^{\Gamma}(x_s)$ since the covariate indices for goods $A$ and $B$ both increase when $(x_s, x_t)\in \mathcal{X}_{s,t}^{1}$. 

Case 2: $(v, \Gamma)\in \mathcal{V}_{1}^{\Gamma}(x_t)$. If $v$ satisfies $\delta_{As}+v_A \geq \delta_{Bs}+v_B$, it implies $(v, \Gamma) \in \mathcal{V}_{1}^{\Gamma}(x_s)$ since the covariate index for good $A$ increases at time $s$ relative to time $t$. Otherwise $v$ should satisfy $\delta_{As}+v_A < \delta_{Bs}+v_B$. Also the complementarity is nonnegative $\Gamma\geq 0$, which implies the following conditions:
\begin{equation*}
\delta_{As}+v_A+\Gamma \geq \delta_{At}+v_A \geq 0, \ \delta_{As}+v_A+\Gamma+ \delta_{Bs}+v_B\geq 2(\delta_{As}+v_A)\geq 0.
\end{equation*}

The above condition implies $(v, \Gamma) \in \mathcal{V}_{2}^{\Gamma}(x_s)\subseteq \mathcal{V}_{D_A}^{\Gamma}(x_s)$. 

In summary, I have shown that $\mathcal{V}_{D_A}^{\Gamma}(x_t) \subseteq \mathcal{V}_{D_A}^{\Gamma}(x_s)$ for  any $(x_s, x_t)\in \mathcal{X}_{s,t}^{1}$, implying that the conditional probability $\Pr(Y_{it}\in D_A\mid \alpha_i, x_s, x_t, \Gamma_{it}\geq 0)$ increases at time $s$ compared to time $t$.
The proof of the lower bound for $\eta$ can be established similarly, so it is omitted here.

\end{proof}

\subsubsection{Proof of Proposition \ref{prop:three}}\label{proof:propthree}

\begin{proof}

%The proof is similar to the proof \ref{proof:prop1} for Proposition \ref{prop1}.
The proof is similar to the proof for Proposition \ref{prop1}.
 We will first derive sufficient conditions for decreased conditional choice probabilities. Then when we observe increased conditional choice probabilities, identifying restrictions on parameters can be derived by contraposition.

 Let  $v_{ij t}=\alpha_{ij}+\epsilon_{ij t}$ denote the sum of fixed effects and error term  for $j \in \mathcal{L}$. Let $v_{j_1j_2 t}=v_{j_1t}+v_{j2t}+\Gamma_{j_1j_2}$ and $v_O=0$ denote the sum of fixed effects and error term for bundle $j_1 j_2$ and the outside option respectively, for $j_1\neq j_2\in \mathcal{L}$.  Given covariate $X_{ijt}=x_{jt}$, let $\delta_{jt}=x_{jt}'\beta_0$, $\delta_{j_1 j_2 t}=\delta_{j_1 t}+\delta_{j_2 t}$, and $\delta_{Ot}=0$.

 For any set $K\subset \mathcal{C}$, let $\mathcal{V}_{K}(x_{t})$ denote the collection of $v=(v_A,..., v_J)$ such that there exists one choice in $K\subset \mathcal{C}$ being chosen given $X_{it}=x_t$. The set $\mathcal{V}_{K}(x_t)$ can be expressed as
 \begin{equation*}
 \mathcal{V}_{K}(x_{t})=\big\{v \mid \exists j\in K \ \text{s.t.} \  \delta_{jt}+v_{j}\geq \delta_{kt}+v_{k} \ \text{for all} \ k \in K^c    \big \},
 \end{equation*}
 
The conditional probability that there exists one choice in set $K$ being chosen can be expressed as follows:
 \begin{equation*}
\Pr(Y_{it}\in K\mid \alpha_i, x_s, x_t)=\Pr\big(v_{it} \in \mathcal{V}_{K}(x_t) \mid \alpha_i, x_s, x_t \big).
 \end{equation*}

 Under Assumption \ref{ass:sta} (stationarity), the conditional distribution of $v_{it}$ is stationarity over time for any $s\neq t$ since the fixed effect $\alpha_i$ is constant over time.  Therefore, a larger set implies a larger conditional choice probability over time: 
\begin{equation*}\label{set}
\mathcal{V}_{K}(x_s) \subseteq \mathcal{V}_{K}(x_t) \ \Longrightarrow \  \Pr(Y_{is}\in K\mid \alpha_i, x_s, x_t) \leq \Pr(Y_{it}\in K\mid \alpha_i, x_s, x_t).  
\end{equation*}

I will show sufficient conditions for the above set inclusion relationship, which implies decreased conditional choice probabilities.  Then identifying restrictions can be derived when increased conditional choice probabilities are observed. Proposition \ref{prop:three} contains three parts of identifying restrictions, and the proof for each part is shown as follows.

\textbf{Part 1}: comparisons of the conditional probability of choosing good $j\in \mathcal{L}$ over time.  I look at good $A$ as an example. The set $\mathcal{V}_{A}(x_{t})$ is given as 
 \begin{equation*}
 \begin{aligned}
 \mathcal{V}_A(x_{t})=\big\{v \mid \delta_{At}+v_{A} \geq 0,  \delta_{At}&+v_{A}-\delta_{j_1 t}-v_{j_1} \geq 0, \delta_{j_1 t}+v_{j_1}+\Gamma_{A j_1} \leq 0, \\
& \delta_{At}+v_{A}-\delta_{j_1 j_2 t}-v_{j_1 j_2}\geq0 \ \text{for all} \ j_1 \neq j_2 \in \{B,..., J\} \big \},
 \end{aligned}
 \end{equation*}
 
 Notice that $\delta_{j_1 j_2 t}=\delta_{j_2 j_1 t} $ and $v_{j_1 j_2}=v_{j_2 j_1}$ by definition, so switching the order of the two indexes of a bundle will not make any difference. 
%when the covariate index of choice $A$ decreases and the covariate index of two other goods increases over time,  i
The set $ \mathcal{V}_A(x_{t})$ increases with respect to $\delta_{At}$ and decreases with $\delta_{kt}$ for $k\in \{B,.., J\}$. Therefore,  the following set relationship holds:
\begin{multline*}
\Delta_{s,t} \delta_{A}\leq 0, \  \Delta_{s,t} \delta_{k} \geq 0 \  \forall  k\in  \{B,.., J\} \ \Longrightarrow  \\
   \Pr(Y_{is}= \{A\} \mid \alpha_i, x_s, x_t) \leq \Pr(Y_{it}=\{A\} \mid \alpha_i, x_s, x_t).
\end{multline*}

By contraposition and taking expectation over $\alpha_i$ conditional on  $(x_s, x_t)$, it yields the first identifying restriction in Proposition \ref{prop:three}:
\begin{equation*}
 P_s(\{A\} \mid x_s, x_t) > P_t(\{A\} \mid x_s, x_t)
 \Longrightarrow \  \{ \Delta_{s,t} \delta_A>0 \} \lor \{ \exists k \in  \{B,.., J\}, \  \Delta_{s,t} \delta_{k}<0\}.
 \end{equation*}

\textbf{Part 2}: comparisons of the conditional probability of choosing bundle $j_1j_2$ over time for $j_1\neq j_2 \in \mathcal{L}$.  I look at bundle $AB$ as an example. The set $\mathcal{V}_{AB}(x_{t})$ is given as 
 \begin{equation*}
 \begin{aligned}
 \mathcal{V}_{AB}(x_{t})=&\big\{v \mid \delta_{ABt}+v_{AB} \geq 0, \  \delta_{\ell t}+v_{\ell }+\Gamma_{AB} \geq 0, \delta_{ABt}+v_{AB}-\delta_{j_1 t}-v_{j_1} \geq 0, \\
& \delta_{AB t}+v_{AB}-\delta_{j_1 j_2 t}-v_{ j_1 j_2} \geq0,  \ \text{for all} \ \ell \in \{A, B\}, j_1\in \{C,..., J\}, j_2\neq j_1 \in \mathcal{L} \big \}.
 \end{aligned}
 \end{equation*}

According to the definition of  $ \mathcal{V}_{AB}(x_{t})$, the set $ \mathcal{V}_{AB}(x_{t})$ increases when the covariate index $\delta_{\ell t}$ increases and  $\delta_{k}$ decreases  for $\ell \in \{A, B\}$ and $k\in \{C,..., J\}$. So the following restrictions hold:%Therefore, when the covariate index of choice $A$ decreases and the covariate index of two other goods increases over time,  it implies the following set relationship:
\begin{equation*}
\begin{aligned}
\Delta_{s,t} \delta_{\ell}\leq 0, \Delta_{s,t} \delta_k  &\geq 0 \  \ \forall  \ell \in \{A, B\},  k\in \{C,..., J\}  \ \Longrightarrow  \\
&  \Pr(Y_{is}=\{AB\} \mid \alpha_i, x_s, x_t) \leq \Pr(Y_{it}=\{AB\} \mid \alpha_i, x_s, x_t).
 \end{aligned}
\end{equation*}

By contraposition and taking expectation over $\alpha_i$ conditional on  $(x_s, x_t)$, it yields the second identifying restriction in Proposition \ref{prop:three}:
\begin{equation*}
\begin{aligned}
 P_s(\{AB\} \mid x_s, x_t) > & P_t(\{AB\} \mid x_s, x_t)
 \Longrightarrow \\
& \big\{ \exists \ell \in \{A, B\}, \ \Delta_{s,t} \delta_{\ell}>0 \big\} \lor    \{ \exists k \in  \{C,..., J\}, \ \Delta_{s,t} \delta_k<0\}.
 \end{aligned}
 \end{equation*}

\textbf{Part 3}:  comparisons of the conditional probability of choosing $D_{\ell}=\{\ell, AB\}$ for  $\ell \in\{A, B\}$ over time.
  I take the set $D_A=\{A, AB\}$ as an example to show the proof. 
  We can express the set $\mathcal{V}_{D_A}(x_t)$ as a union of two sets:  choice $A$ or choice $AB$ has higher utility than other choices not in $D_A$, 
\begin{equation*}
\begin{aligned}
 \mathcal{V}_{D_A}(x_t)=\mathcal{V}_{1}(x_t) \cup \mathcal{V}_{2}(x_t), 
 \end{aligned}
 \end{equation*}
 where $\mathcal{V}_{1}(x_t)$ and $\mathcal{V}_{2}(x_t)$ are defined as
 \begin{equation*}
\begin{aligned}
\mathcal{V}_{1}(x_t)=\big\{v  \mid  &\delta_{At}+v_{A}\geq 0, \  \delta_{At}+v_A  \geq  \delta_{k t}+v_{k}, \  \delta_{j_1 t}+v_{j_1}+\Gamma_{A j_1}\leq 0, \\
& \delta_{At}+v_{A}\geq \delta_{j_1 t}+\delta_{j_2 t}+v_{j_1 j_2}, \ \text{for all} \ k, j_2 \in \{B,..., J\}, j_1\neq j_2\in \{C,..., J\} \},
 \end{aligned}
 \end{equation*}
 and 
  \begin{equation*}
\begin{aligned}
\mathcal{V}_{2}(x_t)=\big\{v  \mid \delta_{ABt}&+v_{AB} \geq 0, \  \delta_{ABt}+v_{AB}  \geq \delta_{k t}+v_k,   \  \delta_{Bt}+v_B+\Gamma_{AB}\geq \delta_{j_1t}+v_{j_1}+\Gamma_{A j_1}, \\ 
&\delta_{ABt}+v_{AB}\geq \delta_{j_1 j_2t}+v_{j_1 j_2} \  \text{for all} \ k, j_2 \in \{B,..., J\}, j_1\neq j_2\in \{C,..., J\} \big\}.
 \end{aligned}
 \end{equation*}
 
We look at the contrapositive statement of the last condition in Proposition \ref{prop:three}:
\begin{equation}\label{r3}
\begin{aligned}
\{ \Delta_{s,t}\delta_{A}\leq 0\} \land \{ \Delta_{s,t} (&\delta_{A}+ \sign(\Gamma_{AB}) \delta_{B}) \leq 0\} \land \{ \Delta_{s, t}(\delta_B-\delta_{k})\leq 0 \} \\ &\land \{ \Delta_{s,t}\delta_k \geq 0\} \ \forall k\in \{C,..,J\}   \Longrightarrow   \mathcal{V}_{D_A}(x_{s})\subseteq  \mathcal{V}_{D_A}(x_{t}).
\end{aligned}
\end{equation}

Now we only need to show the above condition to prove the last restriction in Proposition \ref{prop:three}. 
 The conditions on parameters depend on the sign of the complementarity $\Gamma_{AB}$. 
We show the proof for the case $\Gamma_{AB}\geq 0$, and the idea applies to $\Gamma_{AB} < 0$.

When $\Gamma_{AB}\geq 0$, the conditions on parameters in \eqref{r3} become
\begin{equation}\label{r31}
\{ \Delta_{s,t}\delta_{A}\leq 0\} \land \{ \Delta_{s,t} (\delta_{A}+\delta_{B})\leq 0\} \land \{ \Delta_{s, t}(\delta_B-\delta_{k})\leq 0 \} \land \{ \Delta_{s,t}\delta_{k}\geq 0\} \ \forall k\in \{C,..,J\}.
\end{equation}

Given the above condition, we want to show that any element $v$ belonging to $ \mathcal{V}_{D_A}(x_s)$ also belongs to $ \mathcal{V}_{D_A}(x_t)$ which proves $ \mathcal{V}_{D_A}(x_s)\subset  \mathcal{V}_{D_A}(x_t)$. 
 For any element $v\in \mathcal{V}_{D_A}(x_s)$,  we discuss two cases:  $v \in \mathcal{V}_{2}(x_s)$ and $v \in \mathcal{V}_{1}(x_s) $. 

\textit{Case 1}:  $v \in \mathcal{V}_{2}(x_s)$. By the definition of the set $\mathcal{V}_{2}(x_s)$, this set increases when the indices $(\delta_{As}, \delta_{As}+\delta_{Bs}, \delta_{Bs}-\delta_{ks}, -\delta_{ks})$ for all $k\in \{C,..., J\}$ increase. Therefore, we know that $\mathcal{V}_{2}(x_s)\subset \mathcal{V}_{2}(x_t)$ under condition \eqref{r31}.

\textit{Case 2}: $v \in \mathcal{V}_{1}(x_s) $.  If $v$ satisfies  $ \delta_{At}+v_{A}\geq \delta_{Bt}+v_{B}$ and  $ \delta_{At}+v_{A}\geq \delta_{Bt}+ \delta_{kt}+v_{Bk}$ for all $k\in \{C,..,J\} $, then $v\in \mathcal{V}_{1}(x_t)$ under condition \eqref{r31}. Otherwise $v$ should satisfy one of the following condition:
\[ \{\delta_{Bt}+v_B> \delta_{At}+v_{A}\} \ \text{or} \ \{ \exists k\in \{C,..,J\}, \ \delta_{Bt}+ \delta_{kt}+v_{Bk}>\delta_{At}+v_{A} \}.\]

Since we assume that $\Gamma_{Bk}\leq 0$ for all $k\in \{C,..,J\}$, either of the above conditions would imply 
\[ \delta_{Bt}+v_B>\delta_{At}+v_{A} \geq 0, \ \delta_{Bt}+v_B+\Gamma_{AB}\geq 0. \]

Then we can verify that $v$ satisfies all conditions in the set $ \mathcal{V}_{2}(x_t)$ under condition \eqref{r31},  so that we know $v \in \mathcal{V}_{2}(x_t) \subseteq \mathcal{V}_{D_A}(x_t) $. 

Now we have proved condition \eqref{r31}. By contraposition and taking expectation over $\alpha_i$ leads to the third restriction in Proposition \ref{prop:three}:
\begin{equation*}
\begin{aligned}
 P_s(D_A \mid x_s, x_t) &>  P_t(D_A \mid x_s, x_t) \Longrightarrow  \{ \Delta_{s,t}\delta_{A}> 0\} \lor \{ \Delta_{s,t} (\delta_{A}+ \delta_{B})> 0\} \\
&  \lor \{ \exists k\in \{C,..,J\}, \ \Delta_{s, t}(\delta_B-\delta_{k})> 0 \} \lor \{ \exists k\in \{C,..,J\}, \ \Delta_{s,t}\delta_k < 0\}.
\end{aligned}
\end{equation*}

\end{proof}

\end{document}